\documentclass[useAMS,usenatbib]{mn2e}
\usepackage{natbib}
\usepackage{graphicx}
\usepackage{amssymb}
\usepackage{amsmath}
\usepackage{calc}

\title[Sunspot opacity: ion-atom absorption processes]{The ion-atom absorption processes as one of the factors of the influence on
       the sunspot opacity}

\author[Ignjatovi{\'c} et al.]{Lj. M. Ignjatovi{\' c},$^{1,2}$ A. A. Mihajlov,$^{1,2}$
V. A. Sre{\'c}kovi{\'c}$^{1,2}$ and
M. S. Dimitrijevi{\'c}$^{2,3,4,5}$\\
\\
$^{1}$University of Belgrade,Institute of Physics, P. O. Box 57, 11001 Belgrade, Serbia\\
$^{2}$Isaac Newton Institute of Chile, Yugoslavia Branch, Volgina 7, 11060 Belgrade, Serbia\\
$^{3}$Astronomical Observatory, Volgina 7, 11160 Belgrade 74, Serbia\\
$^{4}$IHIS-Technoexperts, Be\v zanijska 23, 11080 Zemun, Serbia\\
$^{5}$Observatoire de Paris, 92195 Meudon Cedex, France\\
}

\begin{document}

\date{}

\pagerange{\pageref{firstpage}--\pageref{lastpage}} \pubyear{2008}

\maketitle

\label{firstpage}

\begin{abstract}
As a continuation of the previous investigations of the symmetric and strongly
non-symmetric ion-atom absorption processes in the far UV region within the models of
the quiet Sun photosphere, these processes are studied here within a model of the
sunspot. Here we mean the absorption processes in the H$(1s)$+H$^{+}$ and
H$(1s)+X^{+}$ collisions and the processes of the photo-dissociation of the
H$_{2}^{+}$ and H$X^{+}$ molecular ions, where $X$ is one of the metal atoms:
$X=$Na, Ca, Mg, Si and Al. Obtained results show that the influence of the considered
ion-atom absorption processes on the opacity of sunspots in the considered spectral region (110 nm
$\lesssim \lambda \lesssim$ 230 nm) is not less and in some parts even larger than
the influence of the referent electron-atom processes. In such a way, it is shown that
the considered ion-atom absorption processes should be included \emph{ab initio} in the
corresponding models of sunspots of solar-type and near solar-type stars. Apart of that, the spectral
characteristics of the considered non-symmetric ion-atom absorption processes
(including here the case $X$ = Li), which can be used in some further applications,
have been determined and presented within this work.
\end{abstract}


\section{Introduction}

In the previous investigations the significant influence of the relevant ion-atom
absorption processes on the solar photosphere opacity was already demonstrated.
So, in \citet{mih86, mih93, mih94} and \cite{mih07a}) have been studied such
symmetric ion-atom processes, as the molecular ion H$_{2}^{+}$ photo-dissociation
\begin{equation}
\label{eq:sim1} \varepsilon_{\lambda} + \text{H}_{2}^{+} \longrightarrow  \text{H} + \text{H}^{+},
\end{equation}
and the absorption charge exchange in (H$^{+}$ + H)-collisions
\begin{equation}
\label{eq:sim2} \varepsilon_{\lambda} + \text{H}^{+} + \text{H} \longrightarrow \text{H} + \text{H}^{+},
\end{equation}
where H=H(1s), H$_{2}^{+}$ is the hydrogen molecular ion in the ground electronic
state, and $\varepsilon_{\lambda}$ - the energy of a photon with the wavelength
$\lambda$. The significance of these processes was established within the solar
photosphere models from \citet{ver81} and \citet{mal86} in the optical, and from
\citet{ver81} in far UV and EUV regions of $\lambda$. Later, the symmetric
processes (\ref{eq:sim1}) - (\ref{eq:sim2}) were included \emph{ab initio}
in one of the new solar photosphere models \citep{fon09}.

Then, in \citet{mih13} was undertaken the investigation of some non-symmetric
ion-atom absorption processes, namely the photo-dissociation and photo-association
of the the molecular ions
\begin{equation}
\label{eq:nonsim1} \varepsilon_{\lambda} + \text{H}X^{+} \longrightarrow \text{H}^{+} + X,
\end{equation}
\begin{equation}
\label{eq:nonsim3} \varepsilon_{\lambda} + \text{H} + X^{+} \longrightarrow (X\text{H}^{+})^{*},
\end{equation}
and the absorption charge-exchange in the ion-atom collisions
\begin{equation}
\label{eq:nonsim2} \varepsilon_{\lambda}+ \text{H} + X^{+} \longrightarrow \text{H}^{+} + X,
\end{equation}
where $X$ is the ground state atom of one of metals, relevant for the used solar
photosphere model, whose ionization potential $I_{X}$ is smaller than the
hydrogen atom ionization potential $I_{\text{H}}$, $X^{+}$ - the corresponding atomic
ion in its ground state, H$X^{+}$ and ($X\text{H}^{+})^{*}$ - the molecular ion in the
electronic states which are adiabatically correlated (at the infinite internuclear
distance) with the states of the ion-atom systems H + $X^{+}$ and H$^{+}$ + $X$
respectively. These processes were examined within the same solar photosphere model
as in \citet{mih07a}, i.e. the model C from \citet{ver81}, with $X$ = Mg, Si and Al.
Also, in accordance with the composition of the solar atmosphere, the processes of
the type (\ref{eq:nonsim1}) - (\ref{eq:nonsim2}), but with atom He(1s$^{2}$) and ion
H$^{+}$ instead H and $X^{+}$, were included in the consideration. However, it was
established that in the case of this atmosphere (for the difference of some helium
reach stellar atmospheres considered in \citet{ign14}) such processes can be practically neglected.
Since the ion-atom systems with the mentioned $X$ are strongly non-symmetric, the
examined processes generate the quasi-molecular absorption bands in the
neighborhoods of $\lambda$ which correspond to the energies $\Delta_{\text{H};X} \equiv
I_{\text{H}}-I_{X}$. According to the values of $\Delta_{\text{H};X}$ with the mentioned X these
absorption bands lie in the part of the far UV region. We should note here that the
photo-dissociation of the molecular ion HSi$^{+}$ was considered first time from the
astrophysical aspect (interstellar clouds and the atmospheres of red giant stars)
in \citet{sta97}.

In \citet{mih13} was shown that, in the case of the quiet Sun, ion-atom processes
(\ref{eq:sim1}) - (\ref{eq:sim2}) and (\ref{eq:nonsim1}) - (\ref{eq:nonsim2})
together become seriously concurrent to some other relevant absorbtion processes in
far UV and EUV regions within the whole solar photosphere. This result is especially
important, since among all possible ion-atom non-symmetric processes only those of
them, for which the needed data about the corresponding
molecular ions were known, were taken into account. Because of that, it was natural to conclude that the
non-symmetric processes (\ref{eq:nonsim1}) - (\ref{eq:nonsim2}) should be also
included \emph{ab initio} in the corresponding solar photosphere models.

One can see that the previous investigations of the ion-atom absorption
processes were performed only in the case of the quiet Sun. However, it is well
known how significant role for the solar atmosphere the sunspots play, and certainly
it was interesting to see what is the situation with these processes in such
objects. Because of that, this investigation, whose some preliminary
results were referred recently on a corresponding astrophysical conference
\citet{sre13}, was undertaken. All considerations were within the sunspot model M
from \citet{mal86}. Such choice was caused by the fact that only this model, among
other models mentioned in the literature (see e.g. \citet{fon06}), provided all
data needed for the calculations of the absorption coefficients which characterize
the considered absorption processes. Certainly, the ion-atom absorption processes of
the type (\ref{eq:nonsim1}) - (\ref{eq:nonsim2}) were included in the
considerations. However, here we take into account also additional processes of the
molecular ion photo-dissociation and photo-association
\begin{equation}
\label{eq:sat1} \varepsilon_{\lambda} + \text{H}X^{+} \longrightarrow \text{H} + X^{+*},
\end{equation}
\begin{equation}
\label{eq:sat3} \varepsilon_{\lambda} + \text{H} + X^{+} \longrightarrow \text{H}X^{+*},
\end{equation}
where $X^{+*}$ is the ion in excited state with the excitation energy
$E_{exc}(X^{+*}) \lesssim \Delta_{\text{H};X}$ and H$X^{+*}$ - the molecular ion in the
electronic state adiabatically correlated with the state of the H + $X^{+*}$
system, as well as the corresponding collisional excitation
\begin{equation}
\label{eq:sat2} \varepsilon_{\lambda}+ \text{H} + X^{+} \longrightarrow \text{H} +X^{+*},
\end{equation}
which realizes over creation of the quasimolecular complex (H + $X^{+})^{*}$ in the
excited electronic state adiabatically correlated with the state of the same system
H + $X^{+*}$ as in the process (\ref{eq:sat3}). Depending of the values of
$E_{exc}(X^{+*})$ the absorption bands generated by the processes (\ref{eq:sat1}) -
(\ref{eq:sat2}) can lie not only in far UV region of $\lambda$, but also in near UV
and visible regions. The reason for the consideration of these processes is the fact
that the absorption bands generated by some of them overlap with the bands generated
by the processes (\ref{eq:sim1}) - (\ref{eq:sim2}) and (\ref{eq:nonsim1}) -
(\ref{eq:nonsim2}). If the radiative transition $X^{+} \rightarrow X^{+*}$ is
allowed by the dipole selection rules the corresponding absorption band can be
treated, as in the case of the similar phenomena caused by atom-atom collisions
(\citet{vez98,ske02}), i.e. as the satellite of the ion spectral line connected with
the mentioned transition.

The basic task of this investigation is to estimate the significance of the
symmetric and non-symmetric ion-atom absorption processes in the case of the sunspot
with respect to the processes of the negative hydrogen ion H$^{-}$ photo-detachment
and inverse "bremsstrahlung" in ($e$ + H)-collisions, namely
\begin{equation}
\label{eq:eat1} \varepsilon_{\lambda} + \text{H}^{-} \longrightarrow \text{H} + e',
\end{equation}
\begin{equation}
\label{eq:eat2} \varepsilon_{\lambda} + e + \text{H} \longrightarrow \text{H} + e',
\end{equation}
where $e$ and $e'$ denote the free electron in initial and final channel, which,
similarly to the previous papers, are treated here as the referent processes. It is connected with the
concept of this paper which stays the same as in all previous papers. Namely, the
aim of these papers (see e.g. \citet{mih93,mih07a,mih13}) was to pay attention on
the considered ion-atom radiative processes as the factors of the influence on the
opacity of the solar atmosphere. For that purpose it was needed (and enough in the
same time) to show that the efficiency of these processes in the considered spectral
region is close to the efficiency of some known radiative processes whose
significance for the solar atmosphere is accepted in literature. It is clear that
in the case of this atmosphere just the electron-atom processes (\ref{eq:eat1}) and
(\ref{eq:eat2}) can be taken as the referent ones. Because of that in these previous
papers many other radiative processes have not been considered, including here the certainly
very important processes of the metal atom photo-ionization, which were already
discussed in the literature in connection with the quiet Sun atmosphere in
\citet{fon11}. We mean the processes
\begin{equation}
\label{eq:phion} \varepsilon_{\lambda} + (X)_{g}^{*} \longrightarrow X^{+} + e',
\end{equation}
where $(X)_{g}^{*}$ denotes the given metal atom in the ground state, i.e. $X$,
or in any possible (under the considered conditions) excited state, i.e. $X^{*}$.
However, within the sunspot we have a significantly smaller temperature than in the
quiet Sun photosphere (see figure about model) and consequently it is possible to
expect there more larger efficiency of these photo-ionization processes, so that the position of the
mentioned electron-atom processes as the referent
ones is not so clear. Because of that the processes of the metal atom photo-ionization were taken
into account from the beginning of this investigation (some of them were considered
already in \citet{sre13}).
\begin{figure*}
\centering
\includegraphics[height=0.34\textwidth]{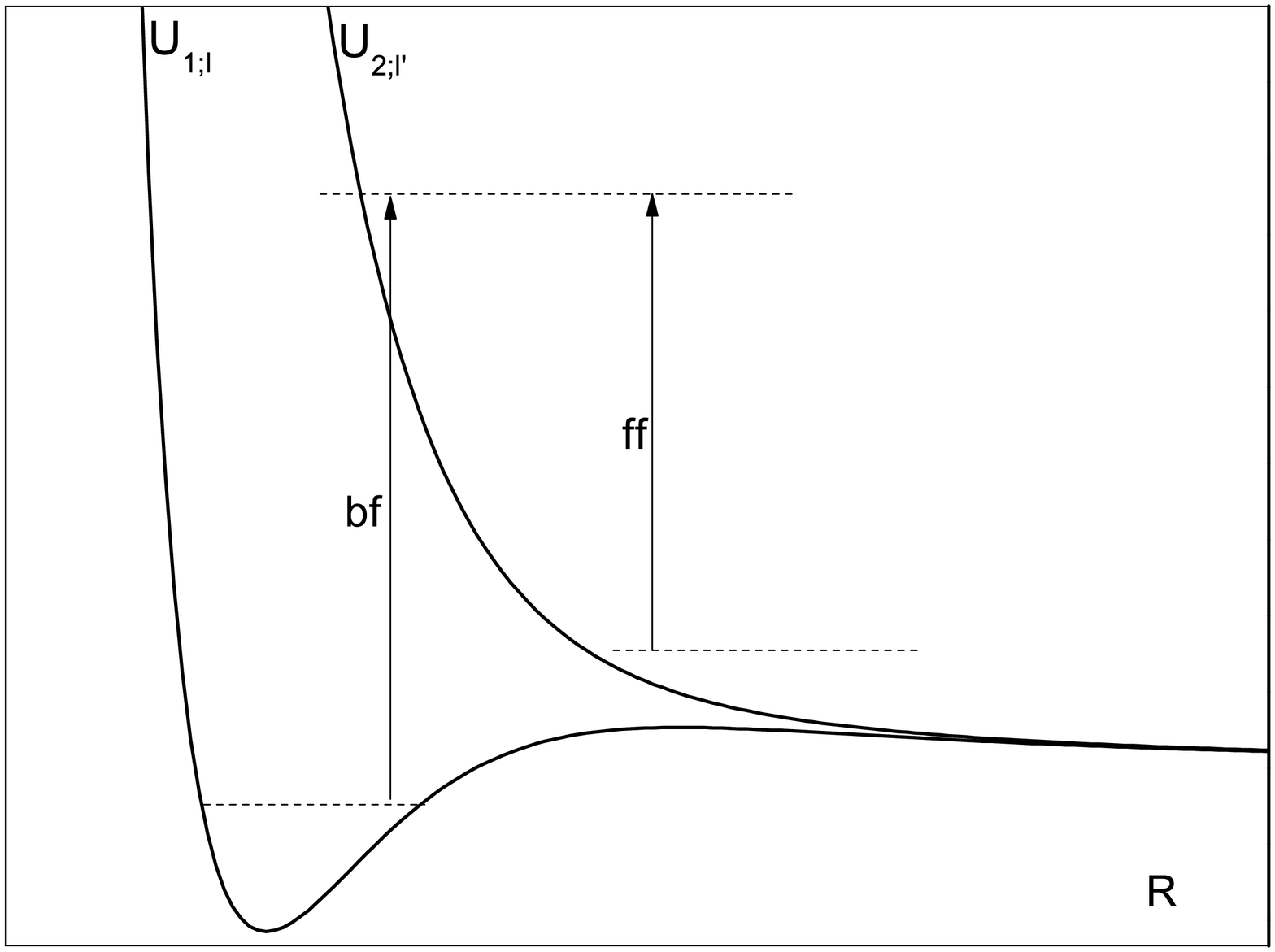}
\includegraphics[height=0.34\textwidth]{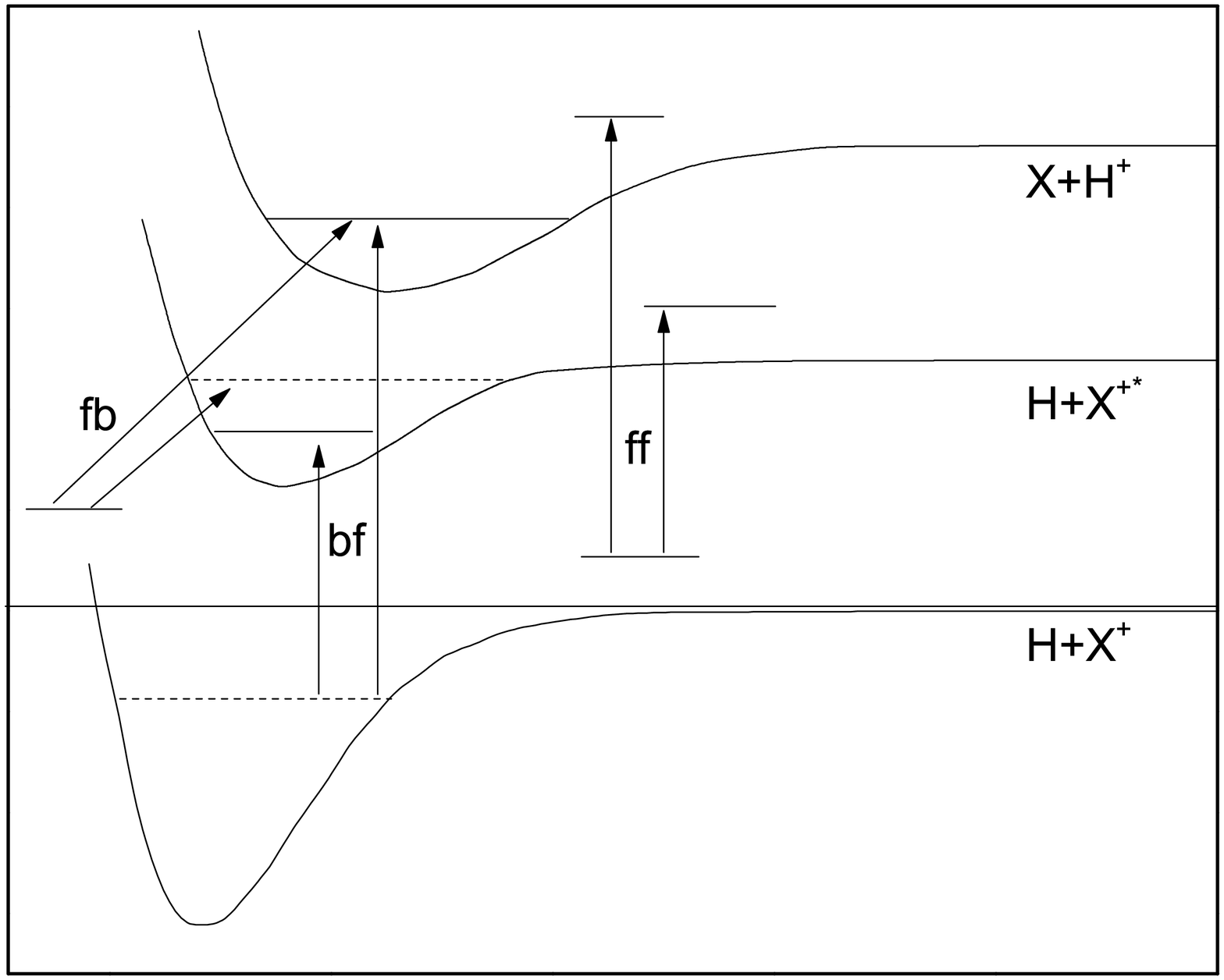}
\caption{\textit{Left panel \bf{a}:}: The bound-free (bf) and free-free (ff)
transitions in the case of the symmetric ion-atom processes (\ref{eq:sim1})-
(\ref{eq:sim2}).
\textit{Right panel \bf{b}:} The bound-free (bf), free-free (ff) and free-bound
(fb) transitions in the case of the non-symmetric ion-atom processes:
$\longrightarrow$ - Eqs. (\ref{eq:nonsim1})- (\ref{eq:nonsim2}),
$-->$ Eqs. (\ref{eq:sat1})-(\ref{eq:sat2}).}
\label{fig:schem}
\end{figure*}
In accordance with above mentioned the following absorption processes are included
in the consideration in this work:\\
- the symmetric ion-atom processes (\ref{eq:sim1}) - (\ref{eq:sim2});\\
- the non-symmetric processes (\ref{eq:nonsim1}) - (\ref{eq:nonsim2}) with $X$ = Na,
  Ca, Mg, Si and Al, for which the needed data about the corresponding molecular ions
  H$X^{+}$ are known;\\
- the non-symmetric processes (\ref{eq:sat1}) - (\ref{eq:sat2}) with $X^{+*}$ =
  Ca$^+(3p^{6}3d)$, Ca$^+(3p^{6}4p)$, Ca$^+(3p^{6}5s)$, Ca$^+(3p^{6}4d)$,
Ca$^+(3p^{6}5p)$ and Al$^+(2p^{6}3s3p)$,\\
- the electron-atom processes (\ref{eq:eat1}) and (\ref{eq:eat2});\\
- the photo-ionization processes (\ref{eq:phion}) with all metal atoms relevant for
  the used sunspots model, i.e. including the case $X$ = Fe.\\
Let us note that as in \citet{sre13} only such non-symmetric processes are taken
into account here for which all needed data about the corresponding molecular ions
are known from the literature.

It is well known that inside sunspot is needed to take into account the presence
of its magnetic field. In connection with this we have to note that according to
the existing data (see e.g. \citet{pen11}) this field is always not larger than 4000
Gs. This is very important for us since it can be shown that in our further
considerations the presence of such magnetic field can be completely neglected and
all needed calculations can be performed as in the case of the quiet Sun.
\begin{figure*}
\centering
\includegraphics[height=0.34\textwidth]{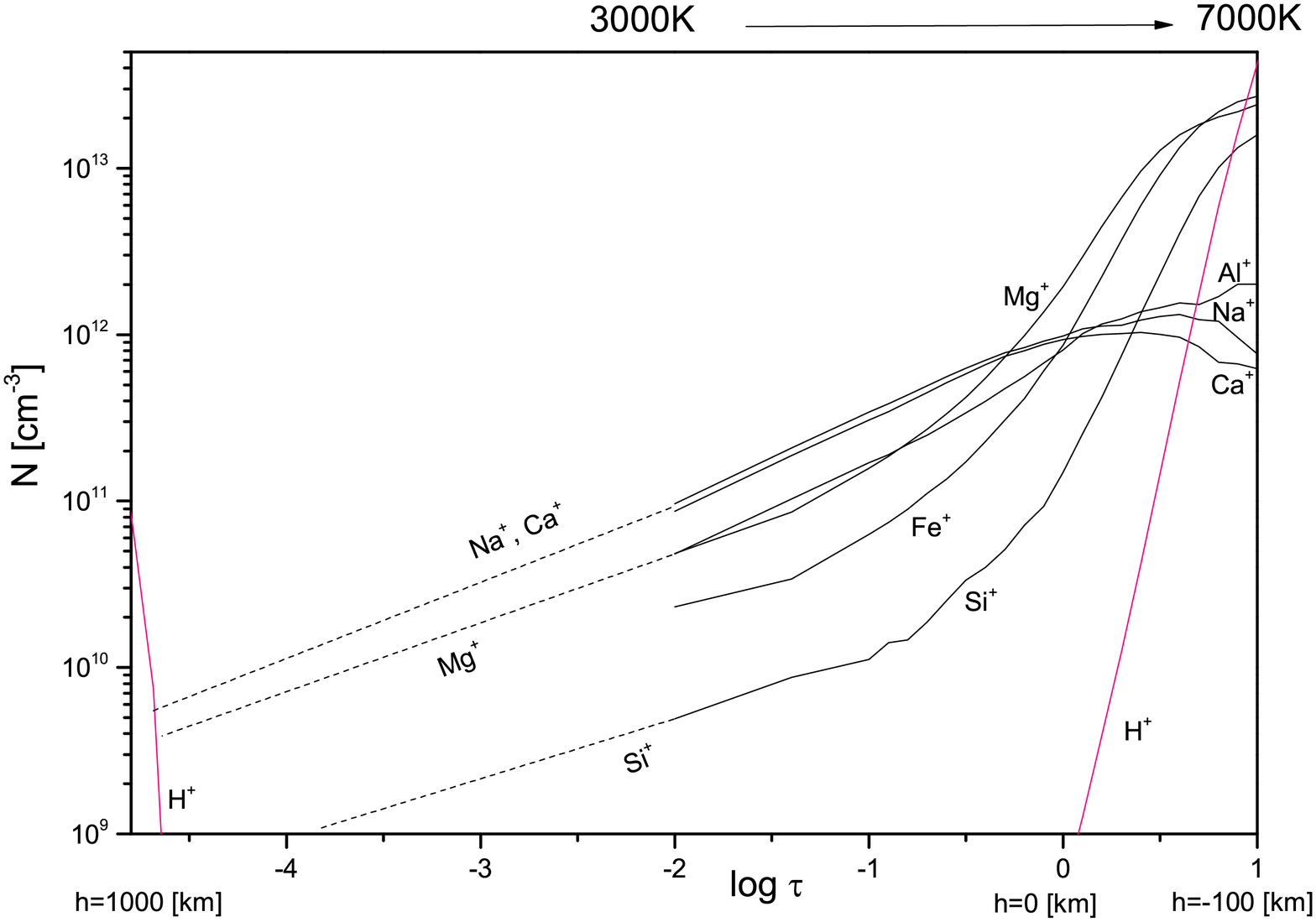}
\includegraphics[height=0.34\textwidth]{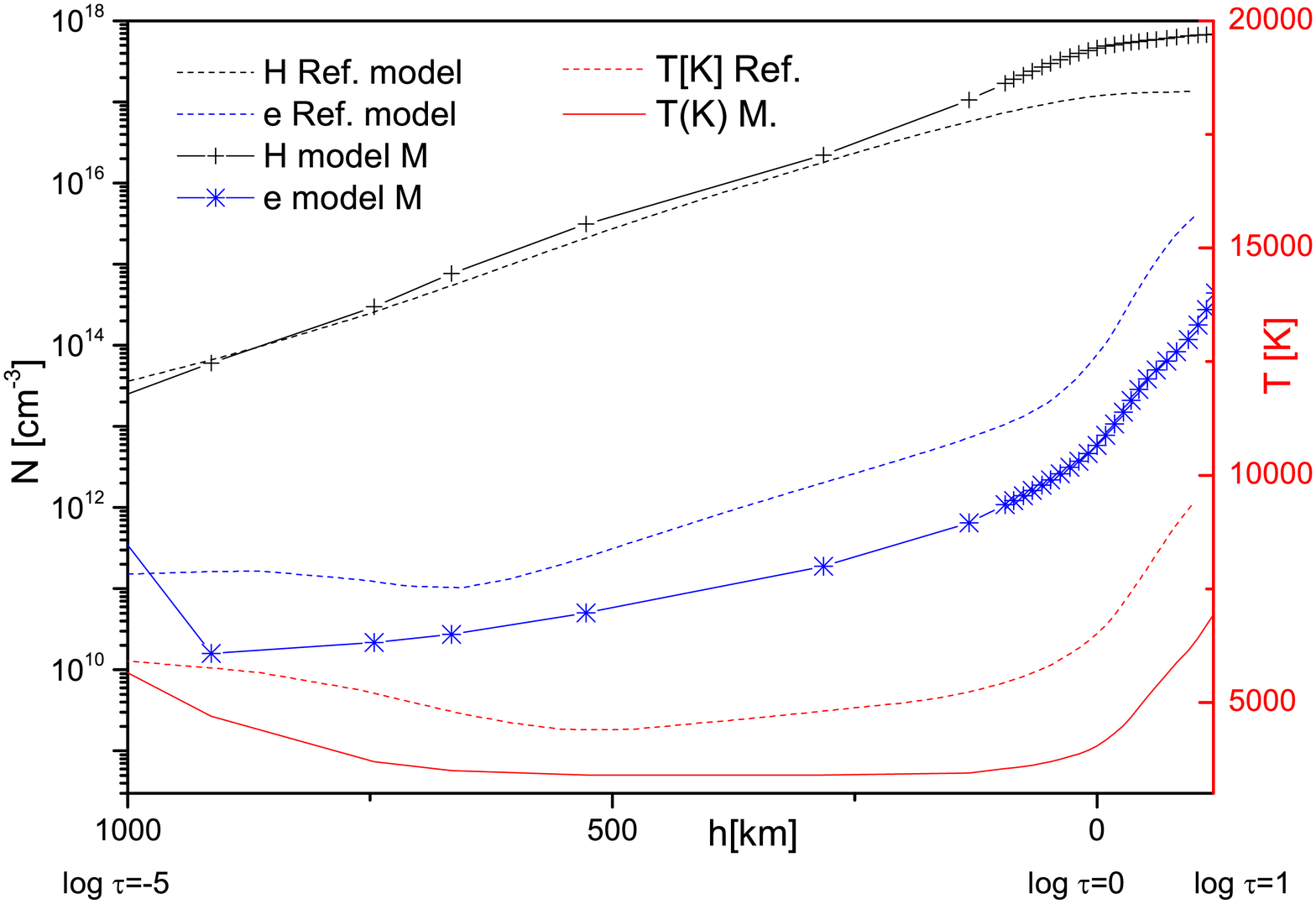}
\caption{\textit{Left panel \bf{a}:}: The hydrogen and metal ion densities
$N_{\textrm{H}^+}$ and $N_{X^+}$ for the sunspot umbral model M from \citet{mal86}.
\textit{Right panel \bf{b}:} The local temperature $T$ and the densities $N(e)$ and
$N(\text{H})$ of the free electrons and hydrogen atoms for the sunspot model M and referent
model of the quiet Sun atmosphere from \citet{mal86}.}
\label{fig:Abund}
\end{figure*}

The aims of this work request determination of the corresponding spectral absorption
coefficients for all mentioned ion-atom processes, as well as for the concurrent
absorption processes (\ref{eq:eat1}) - (\ref{eq:eat2}) and (\ref{eq:phion}), as
functions of $\lambda$ and the height $h$ above the referent solar atmosphere layer.
In this context the processes (\ref{eq:sim1}) - (\ref{eq:sim2}), (\ref{eq:nonsim1})
- (\ref{eq:nonsim2}) and (\ref{eq:sat1}) - (\ref{eq:sat2}) are treated as the
processes from the groups "1", "2" and "3" respectively, which is denotes by the
corresponding index: $j$ = 1, 2 or 3. Let us note that here the spectral
characteristics of the non-symmetric processes of the type (\ref{eq:nonsim1}) -
(\ref{eq:nonsim3}), but with $X$ = Li, are also determined in rather wide regions of
the temperatures and wavelengths. Namely, such processes could be of interest for
lithium rich stellar atmospheres ("Li stars", \citet{hac97, sha01, sha03}) as an
additional canal for the creation of the neutral lithium atoms.

All relevant matter is distributed below in five Sections and two Appendices. So,
the expressions of the considered absorbtion processes are given in the Sections 2
and 3, together with the needed comments about the methods of their determination.
The Section 4 contains the needed comments about used calculation methods. The
results of the calculations of the spectral coefficients and other quantities,
which characterize the relative efficiencies of the considered processes, are
presented (with the corresponding discussions) in the Section 4, and in the last
Section 5 are given some conclusions and are indicated directions of the further
investigations. Then, the potential curves and dipole matrix elements of the
molecular ions H$X^{+}$ with $X$ = Na and Li, which are determined within this work,
are given in Appendix A. Finally, some of the spectral characteristics of the
processes (\ref{eq:nonsim1}) - (\ref{eq:nonsim2}) and (\ref{eq:sat1}) -
(\ref{eq:sat2}), which can be used in some other applications, are presented in
Appendix B.

\section{The spectral characteristics of the ion-atom absorption processes}

\subsection{The partial ion-atom absorption coefficients}

The above mentioned ion-atom absorption processes, namely:\\
- the photo-dissociation (bound-free) processes (\ref{eq:sim1}), (\ref{eq:nonsim1})
  and (\ref{eq:sat1}),\\
- photo-association (free-bound) processes (\ref{eq:nonsim3}) and (\ref{eq:sat3}),\\
- absorption charge-exchange and collisional excitation (free-free) processes
  (\ref{eq:sim2}), (\ref{eq:nonsim2}) and (\ref{eq:sat2})\\
are illustrated by Fig. \ref{fig:schem}.
The efficiencies of these processes are characterized in this work by the corresponding
partial spectral absorption coefficients denoted with:\\
- $\kappa^{(bf)}_{ia;1}(\lambda;h)$, $\kappa^{(bf)}_{iaX;2}(\lambda;h)$ and
  $\kappa^{(bf)}_{iaX;3}(\lambda;h)$, \\
- $\kappa^{(fb)}_{iaX;2}(\lambda;h)$ and $\kappa^{(fb)}_{iaX;3}(\lambda;h),$
- $\kappa^{(ff)}_{ia;1}(\lambda;h)$, $\kappa^{(ff)}_{iaX;2}(\lambda;h)$ and
  $\kappa^{(ff)}_{iaX;3}(\lambda;h)$ respectively.

In accordance with the previous papers these coefficients are given by the similar
expressions, namely
\begin{equation}
\label{eq:kapabf}
\kappa_{ia;1}^{(bf,ff)}(\lambda;h)=K^{(bf,ff)}_{ia;1}(\lambda,T)\cdot
N_{\textrm{H}}N_{\textrm{H}^{+}},
\end{equation}
\begin{equation}
\label{eq:kapaff}
\kappa_{iaX;j}^{(bf,ff)}(\lambda;h)=K^{(bf,ff)}_{iaX;j}(\lambda,T)\cdot
N_{\textrm{H}}N_{X^{+}}, \quad j = 2, 3,
\end{equation}
\begin{equation}
\label{eq:kapafb}
\kappa_{iaX;j}^{(fb)}(\lambda;h)=K^{(fb)}_{iaX;j}(\lambda,T)\cdot
N_{\textrm{H}}N_{X^{+}}, \quad j = 2, 3,
\end{equation}
where $K^{(bf,ff)}_{ia;1}(\lambda,T)$, $K^{(bf,ff)}_{iaX;j}(\lambda,T)$ and
$K^{(fb)}_{iaX;j}(\lambda,T)$, where $j$ = 2 and 3, are the corresponding rate
coefficients. With $T$, $N_{\textrm{H}}$, $N_{\textrm{H}^{+}}$ and
$N_{X^{+}}$ are denoted here the local temperature and densities of the
hydrogen atoms and ions H$^{+}$ and $X^{+}$ respectively, whose values are taken
(for each $h$) from the used sunspot model of \citet{mal86}. Here, as in the
previous papers, it is understood that the photo-dissociation rate coefficients
are defined by the known relations
\begin{equation}
\label{eq:nonumber}
K^{(bf)}_{ia;1}(\lambda,T)=\sigma^{(phd)}_{\text{H}_{2}^{+}}(\lambda,T)
\cdot \chi^{-1}(T;\text{H}_{2}^{+}),
\end{equation}
\begin{equation}
\label{eq:Kbfphd}
K^{(bf)}_{iaX;j}(\lambda,T)=\sigma^{(phd)}_{\text{H}X^{+};j}(\lambda,T)
\cdot \chi^{-1}(T;\text{H}X^{+}), j = 2, 3,
\end{equation}
\begin{equation}
\label{eq:nonumber}
\chi(T;\text{H}_{2}^{+})=\left [\frac{N(\text{H})N(\text{H}^{+})}{N(\text{H}_{2}^{+})} \right]_{T},
\end{equation}
\begin{equation}
\label{eq:Kiaa}
\chi(T;\text{H}X^{+})=\left [\frac{N(\text{H})N(X^{+})}{N(\text{H}X^{+})} \right]_{T},
\end{equation}
where $\sigma^{(phd)}_{\text{H}_{2}^{+}}(\lambda,T)$ and
$\sigma^{(phd)}_{\text{H}X^{+};j}(\lambda,T)$, where $j$ = 2 and 3, are the mean thermal
photo-dissociation cross-sections: for the molecular ions H$_{2}^{+}$ in the process
(\ref{eq:sim1}), and for ions H$X^{+}$ in the processes (\ref{eq:nonsim1}) and
(\ref{eq:sat1}) respectively. With $N_{\text{H}_{2}^{+}}$ and $N_{\text{H}X^{+}}$ are denoted the
densities of the molecular ions $\text{H}_{2}^{+}$ and $\text{H}X^{+}$, and the designation
$[...]_{T}$ denotes that the factors $\chi(T;\text{H}_{2}^{+})$ and $\chi(T;\text{H}X^{+})$ are
determined under the condition of the local thermodynamical equilibrium (LTE) with
given $T$.

\subsection{The total ion-atom absorption coefficients}

The total efficiency of the symmetric ion-atom processes (\ref{eq:sim1}) -
(\ref{eq:sim2}), i.e. group "1", is characterized here by the spectral absorption
coefficient $\kappa_{ia;1}(\lambda;h)$, given by
\begin{equation}
\label{eq:kapasimtot}
\begin{split}
\kappa_{ia;1}(\lambda;h) = K_{ia;1}(\lambda,T)\cdot
N(\textrm{H})N(\textrm{H}^{+}), \quad \\
K_{ia;1}(\lambda,T)=K^{(bf)}_{ia;1}(\lambda,T)+K^{(ff)}_{ia;1}(\lambda,T).
\end{split}
\end{equation}
In connection with the non-symmetric processes we will introduce firstly the total
efficiencies of the processes (\ref{eq:nonsim1}) - (\ref{eq:nonsim2}) and
(\ref{eq:sat1}) - (\ref{eq:sat2}), i.e. groups "2" and "3", for given $X$ which are
characterized by the spectral absorption coefficients $\kappa_{X;2}(\lambda;h)$ and
$\kappa_{X;3}(\lambda;h)$. They are given by
\begin{equation}
\label{eq:nonumber}
\kappa_{X;j}(\lambda;h)= K_{X;j}(\lambda,T)\cdot
N(\textrm{H})N(X^{+}),
\end{equation}
\begin{equation}
\label{eq:kapansimX}
K_{X;j}(\lambda,T)=K^{(bf)}_{X;j}(\lambda,T)+ K^{(fb)}_{X;j}(\lambda,T)+
K^{(ff)}_{X;j}(\lambda,T),
\end{equation}
where $j$ = 2 and 3, and $X$ is the metal atom relevant in these cases. Than,
the total efficiencies of the whole groups "2" and "3", i.e. the non-symmetric
processes (\ref{eq:nonsim1}) - (\ref{eq:nonsim2}) and (\ref{eq:sat1}) -
(\ref{eq:sat2}) with all relevant $X$, are characterized by the spectral absorption
coefficients $\kappa_{ia;2}(\lambda;h)$ and $\kappa_{ia;3}(\lambda;h)$ given by
\begin{equation}
\label{eq:kapansim2-3}
\kappa_{ia;j}(\lambda;h)= \Sigma_{(X)_{j}} \kappa_{X;j}(\lambda;h),
\quad j = 2, 3,
\end{equation}
where $(X)_{j}$ denotes that the summation is performed over all metal atoms
relevant for the processes from the corresponding groups. As the consequence,
the efficiency of symmetric and non-symmetric processes (\ref{eq:nonsim1}) -
(\ref{eq:nonsim2}) is characterized by the spectral absorption coefficient
$\kappa_{ia;1-2}(\lambda;h)$ given by
\begin{equation}
\label{eq:kapansimtot}
\kappa_{ia;1-2}(\lambda;h)= \kappa_{ia;1}(\lambda;h) +
\kappa_{ia;2}(\lambda;h),
\end{equation}
and the total efficiency of all considered ion-atom absorption processes - by the
corresponding spectral absorption coefficient $\kappa_{ia}(\lambda;h)$, namely
\begin{equation}
\label{eq:kapaiatot}
\kappa_{ia;1-3}(\lambda;h)= \kappa_{ia;1-2}(\lambda;h)+ \kappa_{ia;3}(\lambda;h),
\end{equation}
where the coefficients $\kappa_{ia;3}(\lambda;T)$ is given by Eq.
(\ref{eq:kapansim2-3}).

\section{The spectral characteristics of the concurrent absorption processes}

The efficiency of the electron-atom absorption processes (\ref{eq:eat1}) and
(\ref{eq:eat2}) are characterized here by the corresponding bound-free (bf) and
free-free (ff) spectral absorption coefficients $\kappa_{ea;bf}(\lambda;h)$ and
$\kappa_{ea;ff}(\lambda;h)$. They are taken in the usual form
\begin{equation}
\label{eq:nonumber}
\kappa_{ea;bf}(\lambda;h)= K_{ea;bf}(\lambda,T)\cdot N(\textrm{H})N(e),
\end{equation}
\begin{equation}
\label{eq:kapaea12}
\kappa_{ea;ff}(\lambda;h)= K_{ea;ff}(\lambda,T)\cdot N(\textrm{H})N(e),
\end{equation}
from where it follows that the efficiency of these electron-atom processes together
is characterized here by the total absorption coefficient $\kappa_{ea}(\lambda;h)$
given by
\begin{equation}
\label{eq:kapaea}
\begin{split}
\kappa_{ea}(\lambda;h)= K_{ea}(\lambda,T)\cdot N(\textrm{H})N_{e},\\
K_{ea}(\lambda,T) = K_{ea;bf}(\lambda,T) + K_{ea;ff}(\lambda,T).
\end{split}
\end{equation}
where $K_{ea;bf}(\lambda,T)$ and $K_{ea;ff}(\lambda,T)$ are the corresponding
partial rate coefficients. The first of them is defined by the known relations
\begin{equation}
\label{eq:kapaea1}
\begin{split}
K_{ea;bf}(\lambda;h)= \sigma^{(phd)}_{\text{H}^{-}}(\lambda)\cdot
\chi^{-1}(T;\text{H}^{-}),\\
\chi(T;\text{H}^{-})=\left [\frac{N(\text{H})N_{e}}{N(\text{H}^{-})} \right],
\end{split}
\end{equation}
where $\sigma^{(phd)}_{\text{H}^{-}}(\lambda)$ is the spectral cross-section for the ion
H$^{-}$ photo-detachment, the designation $[...]_{T}$ - is already defined above,
and $N(\text{H})$ and the free electron density $N(e)$ are taken from the used sunspot
model.

Than, the photo-ionization processes (\ref{eq:phion}), which are connected with the
metal atoms of the given kinds, are characterized here by the corresponding
effective spectral absorption coefficient $\kappa_{phi;[X]}(\lambda;h)$, namely
\begin{equation}
\label{eq:kapaphi}
\kappa_{phi;X}(\lambda;h)= \sigma^{(phi)}_{[X]}(\lambda;T)\cdot N([X];h),
\end{equation}
where $[X]$ marks the kind of the given metal atoms, $N([X];h)$ - the mean local
density of all these atoms, i.e. the atoms in the ground state and in all excited
states atoms which are realized with existing $N_{e}$ and $T$, and
$\sigma^{(phi)}_{[X]}(\lambda;)$ is the corresponding mean (for given $T$)
photo-ionization cross-sections.

\section{The calculation methods}

\subsection{The ion-atom absorption processes}

As it is known the spectral rate coefficients introduced in Eqs. (\ref{eq:kapabf})
- (\ref{eq:kapafb}) are calculated using the data about the relevant molecular
ions characteristics. We mean the potential curves of the initial and final
electronic states of the considered molecular ions and the corresponding
transition dipole moments. The behavior of these characteristics as the functions of
the internuclear distance $R$ was already shown in the cases of the symmetric and
non-symmetric ($X$ = Mg and Si) ion-atom processes in \citet{mih07a} and \citet{mih13}
respectively. Also, it is illustrated here by figure in the  Appendix A in the case $X$
= Na. The potential curves and transition dipole moments needed for the
non-symmetric processes (\ref{eq:nonsim1}) - (\ref{eq:nonsim2}) with $X=$ Na and Li
are determined within this work and presented in Appendix A. For other above
described non-symmetric ion-atom processes they are taken from \citet{mih13},
\citet{aym12}, \citet{ngu11} and \citet{hab11}, and for symmetric processes - from
\citet{mih07a}.

In this work all mentioned ion-atom processes, excluding (\ref{eq:sim2}) and
(\ref{eq:sat2}) with $X^{+*}$ = Ca$^{+}$($3p^{6}4p$), have completely quantum-mechanical
treatment, and the corresponding partial rate coefficients are determined in the
same way as in \citet{mih07a} and \citet{mih13}, where the whole procedure was
described in details. Since the transition dipole moment of the system H + H$^{+}$
unlimitedly increases (with increasing of $R$), the partial rate coefficient for the
free-free process (\ref{eq:sim2}) is determined here semi-classically, as it was
described in \citet{mih94} and \cite{mih07a}. Then, in the case of the
non-symmetric processes (\ref{eq:sat1}) - (\ref{eq:sat2}) with $X^{+*}$ =
Ca$^{+}(3p^{6}4p)$ the radiative transition Ca$^{+}(3p^{6}4s)\rightarrow$
Ca$^{+}(3p^{6}4p)$ is allowed by the dipole selection rules and, consequently,
quasi-molecular band generated by these processes represent a satellite of the
corresponding ion spectral line. In this case the transition dipole moment approaches
(when $R$ increases) to \textit{constant} $\ne 0$. Because of that the rate
coefficient for this free-free process can be determined by now only
semi-classically. However, in this case it is needed to use the semi-classical
procedure which is described in \cite{ign09} in connection with the ion-atom
processes in helium-rich stellar atmospheres. Let us note that the spectral rate
coefficients of the considered non-symmetric absorption processes, including the
processes with $X$ = Li, are presented here in the Appendix B.

\subsection{The concurrent absorption processes}

The electron-atom absorption processes (\ref{eq:eat1}) - (\ref{eq:eat2}) are
described in this work similarly to \citet{mih94} and \cite{mih07a}. Because of
that the photo-detachment cross-section $\sigma^{(phd)}_{\text{H}^{-}}(\lambda)$ and the
inverse "bremsstrahlung" rate coefficient $K_{ea;ff}(\lambda,T)$ introduced in Eqs.
(\ref{eq:kapaea12}) - (\ref{eq:kapaea1}), which cause the efficiencies of these
processes, are determined here using the data presented in \cite{sti70} and
\cite{wis79}.
\begin{figure*}
\begin{center}
\includegraphics[width=1.4\columnwidth,height=0.95\columnwidth]{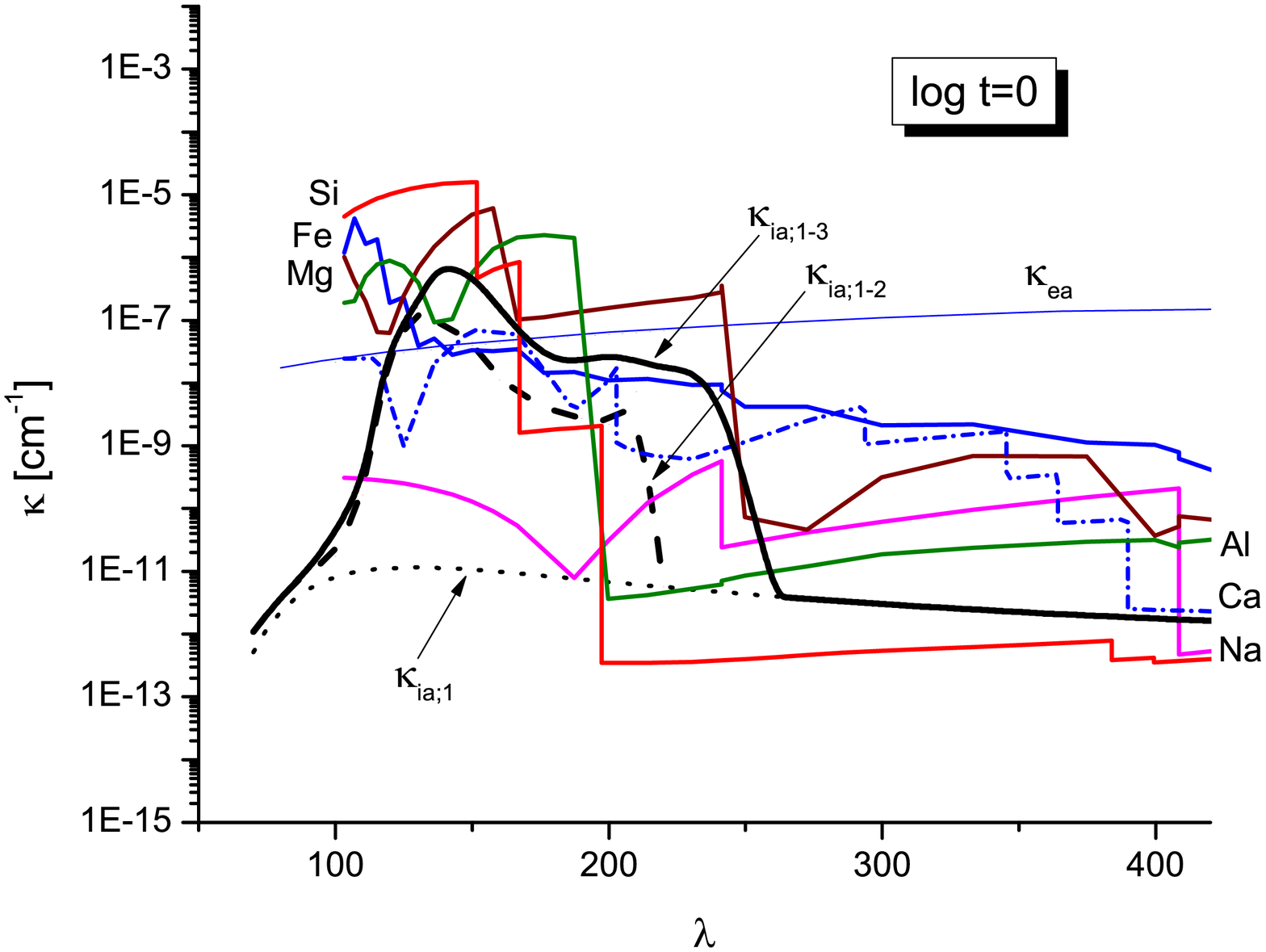}
\caption{The plots of all considered absorbtion processes for log $\tau$ = 0 in the
case of the sunspot (umbral model M from \citet{mal86}): Mg, Si, etc. - the
abbreviations for the spectral coefficients $\kappa_{phi;X}$ of the metal atoms
photo-ionization processes (\ref{eq:phion}) with X = Mg, Si, etc.; $\kappa_{ea}$ -
the electron atom processes (\ref{eq:eat1}) and (\ref{eq:eat2}) together
(H$^{-}$-continuum); $\kappa_{ia;1}$ - symmetric ion-atom processes (\ref{eq:sim1})
and (\ref{eq:sim2}) together (H$_{2}^{+}$-continuum); $\kappa_{ia;1-2}$ - ion-atom
symmetric and non-symmetric processes (\ref{eq:nonsim1}) - (\ref{eq:nonsim2});
$\kappa_{ia;1-3}$ - all ion-atom processes, including non-symmetric processes
(\ref{eq:sat1}) - (\ref{eq:sat2}).}
\label{fig:plot0}
\end{center}
\end{figure*}

According to Eq. (\ref{eq:kapaphi}) the efficiencies of the processes
(\ref{eq:phion}) of the considered metal atoms photo-ionization depend on the
densities $N([X];h)$ of these atoms and on the mean photo-ionization cross-sections
$\sigma^{(phi)}_{[X]}(\lambda;T)$. These cross-sections for the atoms of all
relevant metals (Na, Ca, Mg, Si, Al, Fe) are calculated here in the wide region of
$\lambda$ and $T$ on the basis of the data from \citet{tra68}. The densities
$N([X];h)$ are determined from Saha's equation, with given $T$, $N(e)$ and the ion
densities $N(X^{+})$. For that purpose is needed to know the corresponding
partition functions $Q_{[X]}(T;\delta_{X})$, where $\delta I_{X}\equiv
\delta I_{X}(T;N(e))$ is the lowering of the ionization potential of the given
ground state atom $X$. The calculation of the partition functions, which is the key
element of the whole procedure of $N([X];h)$ determination, is performed here on the
basis of the data from \citet{dra65}.

However, because of this source of the data about the partition functions it is
needed to stay for a moment on this point. The reason for this is the fact that the
partition functions are calculated in \citet{dra65} for the values of $\delta
I_{X}$ from the region which is limited from below with 0.1 eV, while in our case
($N(e)\cong 10^{12} \text{cm}^{-3}, T\cong 3500$ K) we have that $\delta I_{X}\sim 10^{-4}$
eV. As it is known in the Saha's equation $\delta I_{X}$ appears also in the
argument of its exponent, but in the considered case $\delta I_{X}/kT \lesssim
10^{-3}$ and the influence of $\delta I_{X}$ on the values of that exponent can be
completely neglected. However, the fact that existing $\delta I_{X}\ll$ 0.1 eV has
to be examined from the aspect of its influence on the partition functions. Namely,
the moving of the value of $\delta I_{X}$ from 0.1 eV to about 10$^{-4}$ eV causes
the increasing of the principal quantum number of the last realized excited atomic
state from about 11 to about 200. It was shown that such increasing of the largest
principal quantum number causes the increasing of the values of the partition
functions only for about 0.5 percentage. This result means that the partition
functions calculated by using the data from \citet{dra65}, which relate to
$\delta_{X}=$ 0.1 eV, were applicable for determination of the densities $N([X];h)$.

\section{Results and Discussion}

The sunspot model M from \citet{mal86}, as well as its differences with the respect
to the quiet Sun referent model, also from \citet{mal86}, are illustrated by Fig.
\ref{fig:Abund}. So, the part of this figure ("a") presents the densities of
the hydrogen ions H$^{+}$ and H$^{-}$ and relevant metal ions (Na$^{+}$, Ca$^{+}$,
etc.) as the functions of $h$. The part ("b") presents the local temperature and the
densities of the free electrons and hydrogen atoms ($T$, $N_{e}$ and $N(\text{H})$) from
the model M together with the corresponding quantities ($T_{ref}$, $N_{e;ref}$
$N_{ref}(\text{H})$) from the referent model, also as the functions of $h$.

In accordance with the mentioned in the Introduction we should obtain firstly the
picture which reflects the relative efficiencies within the sunspot of all
symmetric and non-symmetric ion-atom processes (\ref{eq:sim1}) - (\ref{eq:sim2}),
(\ref{eq:nonsim1}) - (\ref{eq:nonsim2}) and (\ref{eq:sat1}) - (\ref{eq:sat2}), as
well as the electron-atom absorption processes (\ref{eq:eat1}) - (\ref{eq:eat2})
and metal atom photo-ionization processes (\ref{eq:phion}). For that purpose the
several plots of spectral absorption coefficients of all these processes were
performed for the values of log $\tau$ between $1.0$ and $-1.0$ in the region 70 nm
$\le \lambda \le$ 800 nm. We mean on the ion-atom coefficients
$\kappa_{ia;1}(\lambda;h)$, $\kappa_{ia;1-2}(\lambda;h)$ and
$\kappa_{ia;1-3}(\lambda;h)$, which are given by Eqs. (\ref{eq:kapasimtot}) -
(\ref{eq:kapaiatot}), as well as on the electron-atom and photo-ionization
coefficients, namely $\kappa_{ea}(\lambda;h)$ and  $\kappa_{phi;[X]}(\lambda;h)$,
which are given by (\ref{eq:kapaea}) and (\ref{eq:kapaphi}) respectively. It was
established that in the short wave part of this spectral region, generally we have the
domination of the photo-ionization processes (\ref{eq:phion}) in respect to all
other considered absorption processes. So, for $-1.0\le$ log $\tau \le 0.5$ the
mentioned domination exists in the region $\lambda \lesssim$ 250 nm, but with the
increasing of the value of log$\tau$ this region decreases and for log $\tau = 1.0$
the region of this domination is $\lambda \lesssim$ 175 nm.

\begin{figure*}
\begin{center}
\includegraphics[width=1.4\columnwidth,height=0.93\columnwidth]{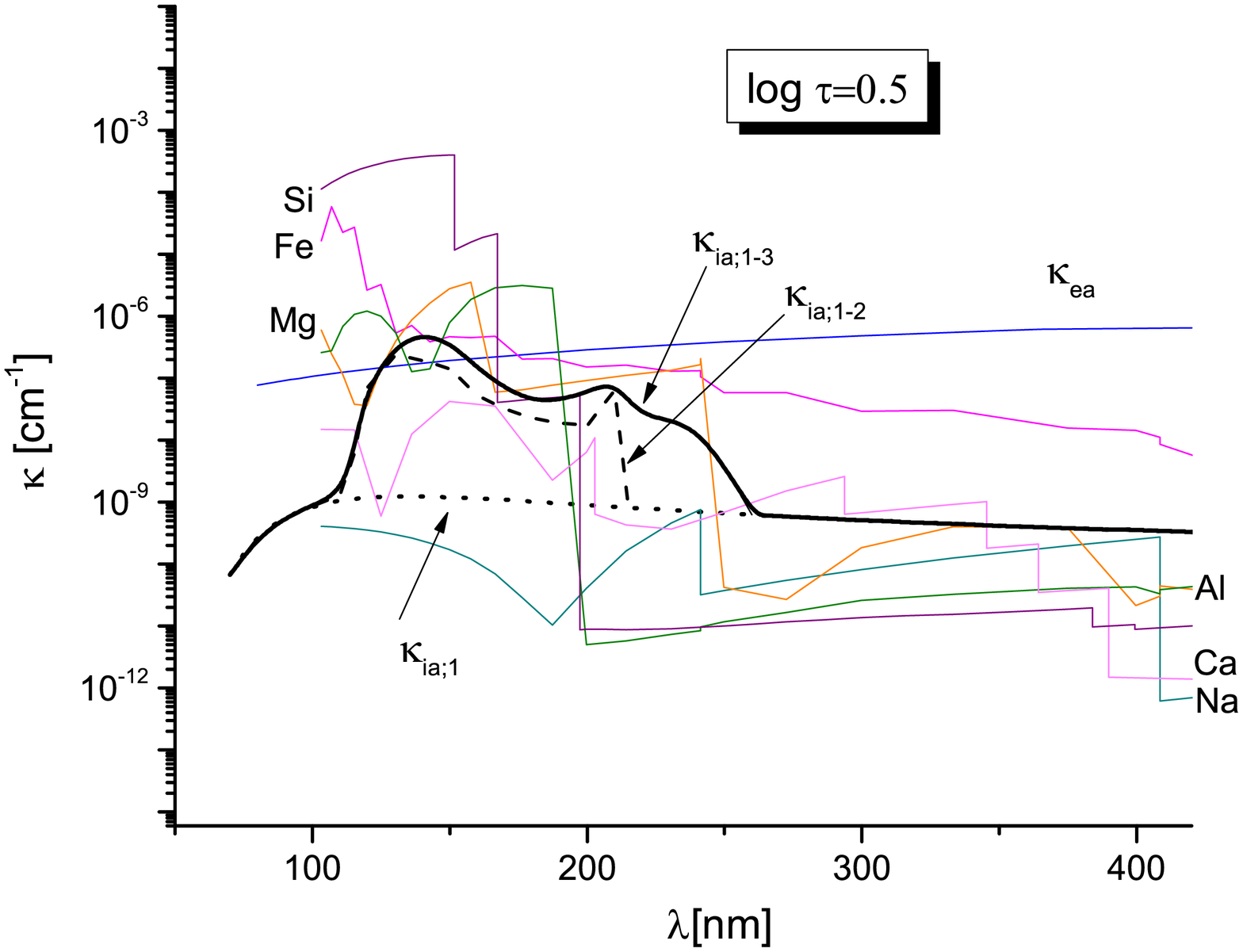} \caption{Same as in the
Fig. \ref{fig:plot1} but for log $\tau$ = 0.5}
\label{fig:plot1}
\end{center}
\end{figure*}

In the same time it was established that the electron-atom absorption processes
(\ref{eq:eat1}) - (\ref{eq:eat2}) can be treated now as the referent processes not
only in the case of the quiet Sun photosphere, but also in the case of a sunspot.
Moreover, for any value of log $\tau$ just these processes dominate in the best part
of the whole considered spectral region in respect to all other absorption processes,
including the processes (\ref{eq:phion}). All mentioned is illustrated by Fig's.
\ref{fig:plot0} and \ref{fig:plot1} which show the plots of all considered
absorption processes for log $\tau$ = 0.5 and 0.0 respectively. Let us note that
the plots for log $\tau$ = 0.5 were needed since just in the region log $\tau >$
0.0 the density of the H$^{+}$ ions becomes very fast increasing.

In Fig's. \ref{fig:plot0} and \ref{fig:plot1} the plot of each of all considered
photo-ionization processes (\ref{eq:phion}) is shown separately. These plots are
marked by designations of the corresponding atoms ("Mg", "Si", etc.). We draw
attention that here was not planed to obtain the plot describing the total efficiency
of all photo-ionization processes. Namely to obtain such a plot would be certainly
useful in the context of the synthesis of the corresponding sunspot spectrum. However,
such task is out of the frame of the present work.

The electron-atom processes (\ref{eq:eat1}) - (\ref{eq:eat2}), which are treated
sometimes as the H$^{-}$-continuum, are represented here by their common
plot which in these figures is marked by $\kappa_{ea}$. However, in accordance with
the aims of this work, the examined ion-atom absorption processes are characterized
by three corresponding plots, namely: one for the whole group "1", i.e. the
symmetric processes (\ref{eq:sim1}) and (\ref{eq:sim2}), which are known also as the
H$_{2}^{+}$-continuum; second for these first processes together with the
non-symmetric processes of the group "3", i.e. (\ref{eq:nonsim1}) -
(\ref{eq:nonsim2}); third for all considered ion-atom processes together, including
the non-symmetric processes of the group "3", i.e. (\ref{eq:sim1}) -
(\ref{eq:sim2}). These plots are marked here by "$\kappa_{ia;1}$",
"$\kappa_{ia;1-2}$" and "$\kappa_{ia;1-3}$" respectively. The presented ion-atom
plots show that it was indeed necessary to take into account additional
non-symmetric processes of the group "3". Namely, one can see that the inclusion of
these processes significantly increases the total
efficiency of the non-symmetric ion-atom absorption processes.

The figures \ref{fig:plot0} and \ref{fig:plot1} demonstrate that the total
efficiencies of all considered symmetric and non-symmetric ion-atom absorption
processes in the region 110 nm $\lesssim \lambda \lesssim$ 230 nm is of the same
order of the magnitude as the one of the electron-atom processes (\ref{eq:eat1}) -
(\ref{eq:eat2}) together (H$^{-}$-continuum). However, here is needed to
emphasis that in the significant part of this region these total efficiencies are
close, or the ion-atom efficiency is even larger than electron-atom one.
\begin{figure}
\begin{center}
\includegraphics[width=\columnwidth,
height=0.75\columnwidth]{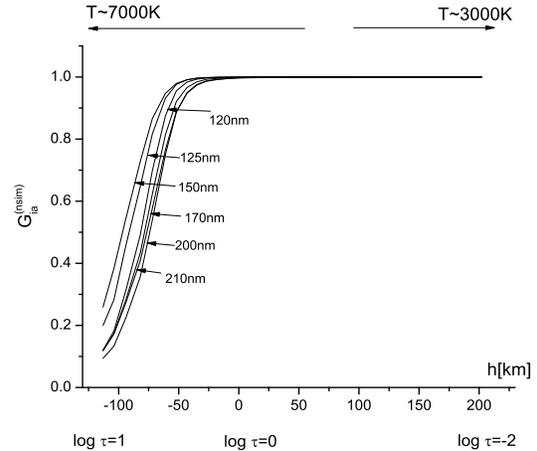} \caption{The behavior of the
quantity $G^{(nsim)}_{ia}(\lambda;h)$, given by equation (\ref{eq:Gnsim}), which
characterizes the relative efficiency of all non-symmetric ion-atom processes in
respect to the total efficiency of all ion-atom absorption precesses
(the sunspot model M from \citet{mal86}).}
\label{fig:Gqs}
\end{center}
\end{figure}

From Fig. \ref{fig:Abund} one can see that the plasma temperature in sunspot ($T$)
is significantly smaller than in the case of the quiet Sun ($T_{ref}$) in the larger
part of that region of log $\tau$ or $h$ which is shown in this figure. Consequently,
it was possible to expect that in this region the relative importance of the
non-symmetric absorption processes (in respect to the symmetric ones), at least for
110 nm $\lesssim \lambda \lesssim$ 230 nm, will be larger than in the case of the quiet
Sun. In order to test this assumption and to establish what is the contribution of
all considered non-symmetric processes to the total ion-atom efficiencies in far UV
region of $\lambda$ we have determined the adequately defined quantity
$G^{(nsim)}_{ia}(\lambda;h)$, namely
\begin{equation}
\label{eq:Gnsim}
G^{(nsim)}_{ia}(\lambda;h) =
\frac{\kappa_{ia;2}(\lambda;h) + \kappa_{ia;3}(\lambda;h)}
{\kappa_{ia;1-3}(\lambda;h)},
\end{equation}
where $\kappa_{ia;j}(\lambda)$, with $j$ = 2 and 3, and $\kappa_{ia}(\lambda;h)$
are given by Eqs. \ref{eq:kapansim2-3} and \ref{eq:kapaiatot} respectively. The
behavior of this quantity (for several values of $\lambda$) is illustrated by  Fig.
\ref{fig:Gqs} in the interval of heights: -125 km $\le h \le$ 200 km. From this
figure one can see that in far UV region the non-symmetric processes are dominant in
respect to the symmetric ones for $h >$ 50 km, i.e. in the largest part of the
region of $h$ shown in Fig. \ref{fig:Abund}, but for -75 km $\le h \lesssim$ 50 km
the efficiencies of the symmetric and non-symmetric processes are close. This fact
shows that in the case of the sunspot (as in the case of the quiet Sun) it is useful
to treat the symmetric and all non-symmetric ion-atom absorption processes together.
\begin{figure*}
\begin{center}
\includegraphics[width=1.4\columnwidth,height=0.95\columnwidth]{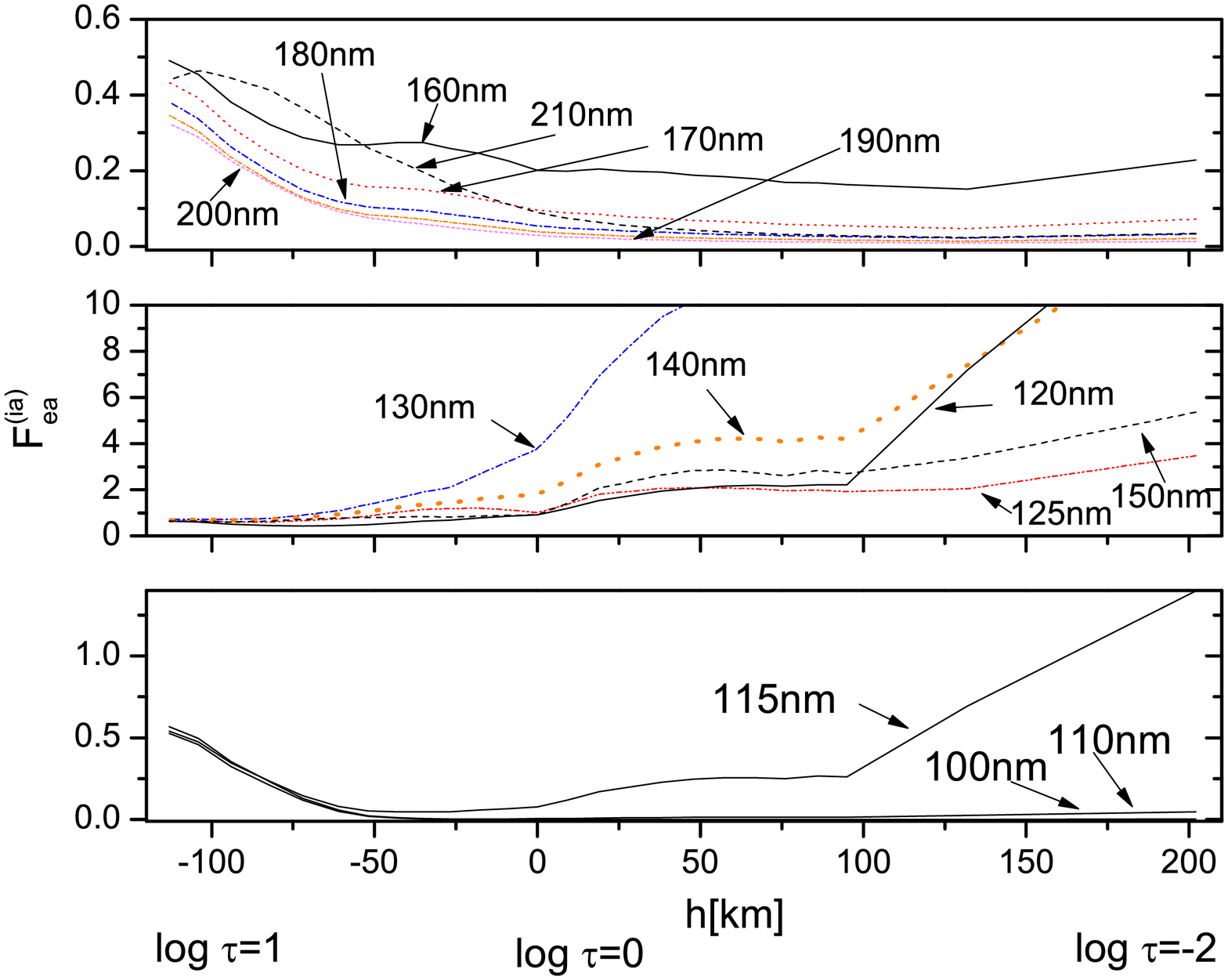}
\caption{The behavior of the quantity $F^{(ia)}_{ea}(\lambda;h)$, given by equation (\ref{eq:Ftotea}),
which characterizes the relative efficiency of all (symmetric and non-symmetric) ion-atom
processes with respect to the total efficiency of the electron-atom absorption
precesses (\ref{eq:eat1}) - (\ref{eq:eat2}) within the sunspot model M from
\citet{mal86}.}
\label{fig:F}
\end{center}
\end{figure*}
Finally, we will have to compare within the same part of the sunspot (which is shown
in Fig. \ref{fig:Abund}) the total efficiencies of the ion-atom and referent
electron-atom absorption processes (\ref{eq:eat1}) and (\ref{eq:eat2}) together
(H$^{-}$-continuum) in the considered part of the far UV region of $\lambda$.
Already this figure gives the possibility to expect that in the sunspot case these
total efficiencies could be close at least in a part of the mentioned spectral
region. It can be tested on the basis of the behavior of the quantity
$F^{(ia)}_{ea}(\lambda;h)$, defined by
\begin{equation}
\label{eq:Ftotea}
F^{(ia)}_{ea}(\lambda) = \frac{\kappa_{ia;1-3}(\lambda;h)}{\kappa_{ea}(\lambda;h)},
\end{equation}
\noindent where $\kappa_{ea}(\lambda;h)$ is given by Eq. \ref{eq:kapaea}. The behavior of
this quantity (also for several values of $\lambda$ from the far UV region) is
illustrated by Fig. \ref{fig:F} in the same interval of heights: -125 km $\le
h\le$ 200 km. This figure shows that: for any of the taken $\lambda$ the total
efficiency of all ion-atom absorption processes is close, or even larger
than the efficiency of the referent electron-atom processes (H$^{-}$-continuum)
in the largest part of the mentioned interval of $h$. This means that in the case
of the sunspot it is needed to consider the discussed ion-atom absorption process
in the mentioned part of the far UV region (110 nm $\lesssim \lambda \lesssim$ 230 nm)
always together with the referent electron-atom processes (H$^{-}$-continuum).
Consequently, the considered ion-atom processes should be also included \emph{ab initio}
in the corresponding models of sunspots of solar-type and near solar-type stars.
Due to possible future investigations of these ion-atom processes, the needed
spectral rate coefficients are determined and presented in Appendix B. Since the
ion-atom processes of the type (\ref{eq:nonsim1}) - (\ref{eq:nonsim2}) with
$X$ = Li could be interesting in connection with the so called lithium stars
(mentioned in the Introduction), the corresponding spectral rate coefficients
are also presented in this Appendix.

\section{Conclusions}

From the presented material it follows that the considered symmetric and
non-symmetric ion-atom absorption processes  influence on
the opacity of sunspots in the considered spectral region (110 nm
$\lesssim \lambda \lesssim$ 230 nm) not less and in some parts even larger than
the referent electron-atom processes. The presented results show that further investigations of the
non-symmetric ion-atom absorption processes promise that their
efficiency could be increased considerably. Namely, the processes (\ref{eq:nonsim1}) -
(\ref{eq:nonsim2}) with $X=$ Fe were not considered here because of the absence of
the data about the needed characteristics of the corresponding molecular ion.
However the Fe component, according to  Fig. \ref{fig:Abund}, gives the
significant contribution to the electron density. Also, some of the possible
processes of the (satellite) type (\ref{eq:sat1}) - (\ref{eq:sat2}) could be very efficient.
It means that the inclusion in the consideration of all possible relevant non-symmetric
ion-atom absorption processes would surely increase their total efficiency.
Because of that, we take such inclusion as the task for the investigations in the
nearest future. For this purpose, here are presented the spectral
characteristics of the considered ion-atom absorption processes which can be used
in some further applications.

Finally, the presented results show that in far UV region the
significance of the considered ion-atom absorption processes is sufficiently large that
they should be included \emph{ab initio} in the corresponding models of the
sunspots. Apart of that, obtained results could be useful also in the case of different
solar like atmospheres.

\section*{Acknowledgments}

The authors wish to thank to Profs. V.N. Obridko, A.A. Nusinov and N. S. Polosukhina
for the shown attention to this work. Also, the authors are thankful
to the Ministry of Education, Science and Technological
Development of the Republic of Serbia for the support of this work within the
projects 176002, III4402.


\begin{thebibliography}{}

\bibitem[Aymar
\& Dulieu(2012)]{aym12} Aymar, M., \& Dulieu, O.\ 2012, Journal of Physics B Atomic
Molecular Physics, 45, 215103

\bibitem[Drawin\& Felenbok(1965)]{dra65} Drawin, H.-W., \& Felenbok, P.\ 1965,
Data for plasmas in local thermodynamic equilibrium, by Drawin, Hans-Werner.;
Felenbok, Paul.~ Paris, Gauthier-Villars, 1965.


\bibitem[Fontenla et al.(2006)]{fon06} Fontenla, J.~M.,
Avrett, E., Thuillier, G., \& Harder, J.\ 2006, ApJ, 639, 441

\bibitem[{{Fontenla} {et~al.}(2009){Fontenla}, {Curdt}, {Haberreiter},
  {Harder}, \& {Tian}}]{fon09}
{Fontenla}, J.~M., {Curdt}, W., {Haberreiter}, M., {Harder}, J., \& {Tian}, H.
  2009, ApJ, 707, 482

\bibitem[Fontenla et al.(2011)]{fon11} Fontenla, J.~M.,
Harder, J., Livingston, W., Snow, M.,
\& Woods, T.\ 2011, Journal of Geophysical Research (Atmospheres), 116, 20108


\bibitem[Habli et al.(2011)]{hab11} Habli, H., Ghalla, H.,
Oujia, B., \& Gad{\'e}a, F.~X.\ 2011, European Physical Journal D, 64, 5

\bibitem[Hack et al.(1997)]{hac97} Hack, M., Polosukhina, N.~S., Malanushenko, V.~P.,
\& Castelli, F.\ 1997, A\& A, 319, 637

\bibitem[Heine(1970)]{hei70} Heine V. \ 1970,
Solid State Physics 24, 1-36.

\bibitem[Ignjatovi{\'c}
\& Mihajlov(2005)]{ign05} Ignjatovi{\'c}, L.~M., \& Mihajlov, A.~A.\ 2005,
Phys.Rev.A, 72, 022715

\bibitem[Ignjatovi{\'c} et al.(2008)]{ign08} Ignjatovi{\'c},
L.~M., Mihajlov, A.~A.,
\& Klyucharev, A.~N.\ 2008, Journal of Physics B Atomic Molecular Physics, 41,
025203

\bibitem[{{Ignjatovi{\'c}} {et~al.}(2009){Ignjatovi{\'c}}, {Mihajlov}, {Sakan},
{Dimitrijevi{\'c}}, \& {Metropoulos}}]{ign09}
{Ignjatovi{\'c}}, L.~M., {Mihajlov}, A.~A., {Sakan}, N.~M., {Dimitrijevi{\'c}},
M.~S., \& {Metropoulos}, A. 2009, MNRAS, 396, 2201

\bibitem[{{Ignjatovi{\'c}} {et~al.}(2014){Ignjatovi{\'c}}, {Mihajlov}, {Sre\'ckovi\'c},
\& {Dimitrijevi{\'c}} }]{ign14}
{Ignjatovi{\'c}}, L.~M., {Mihajlov}, A.~A., {Sre\'ckovi\'c}, V.~A., \&
{Dimitrijevi{\'c}}, M.~S. 2014, MNRAS, 439 (3), 2342


\bibitem[Maltby et al.(1986)]{mal86} Maltby, P., Avrett,
E.~H., Carlsson, M., Kjeldseth-Moe, O., Kurucz, R. L., Loeser, R.\ 1986, ApJ, 306, 284

\bibitem[{Mihajlov \& Dimitrijevi{\' c}(1986)}]{mih86}
Mihajlov, A.~A., \& Dimitrijevi{\' c}, M.~S. 1986, A\&A, 155, 319

\bibitem[{Mihajlov {et~al.}(1993)Mihajlov, Dimitrijevi{\' c}, \& Ignjatovi{\'
  c}}]{mih93}
Mihajlov, A.~A., Dimitrijevi{\' c}, M.~S., \& Ignjatovi{\' c}, L.~M. 1993,
  A\&A, 276, 187

\bibitem[{Mihajlov {et~al.}(1994)Mihajlov, Dimitrijevi{\' c},
  Ignjatovi{\' c}, \& Djuri{\' c}}]{mih94}
Mihajlov, A.~A., Dimitrijevi{\' c}, M.~S., Ignjatovi{\' c}, L.~M., \& Djuri{\'
  c}, Z. 1994, A\&AS, 103, 57

\bibitem[{{Mihajlov} \& {Ignjatovi{\'c}}(1996)}]{mih96}
{Mihajlov}, A.~A., \& {Ignjatovi{\'c}}, L.~M. 1996, DIAM Dynamique des Ions,
  des Atomes et des Molecules, Contrib. Papers, p. 157

\bibitem[{Mihajlov {et~al.}(2007)Mihajlov, Ignjatovi\'c, Sakan, \&
  Dimitrijevi\'c}]{mih07a}
Mihajlov, A.~A., Ignjatovi\'c, L.~M., Sakan, N.~M., \& Dimitrijevi\'c, M.~S.
  2007, A\&A, 437, 1023

\bibitem[{Mihajlov {et~al.}(2013)Mihajlov, Ignjatovi\'c, Sre\'ckovi\'c,
  Dimitrijevi\'c \&bMetropoulos}]{mih13}
Mihajlov, A.~A., Ignjatovi\'c, L.~M., Sre\'ckovi\'c, V.~A., Dimitrijevi\'c, M.~S. \&
Metropoulos, A.  2013, MNRAS, 431, 589

\bibitem[Nguyen et al.(2011)]{ngu11} Nguyen, J.~H.~V.,
Viteri, C.~R., Hohenstein, E.~G., Sherill, C.~D., Brown, K.~P., \& Odom, B. 2011, New Journal of Physics, 13,
063023

\bibitem[Penn
\& Livingston(2011)]{pen11} Penn, M.~J., \& Livingston, W.\ 2011, IAU Symposium,
273, 126

\bibitem[Shavrina et
al.(2001)]{sha01} Shavrina, A.~V., Polosukhina, N.~S., Zverko, J., et al.\ 2001, A\& A, 372, 571

\bibitem[Shavrina et
al.(2003)]{sha03} Shavrina, A.~V., Polosukhina, N.~S., Pavlenko, Y.~V., et al.\
2003, A\& A, 409, 707

\bibitem[{{Skenderovi{\'c}} {et~al.}(2002){Skenderovi{\'c}}, {Beuc}, {Ban}, \&
  {Pichler}}]{ske02}
{Skenderovi{\'c}}, H., {Beuc}, R., {Ban}, T., \& {Pichler}, G. 2002, European
  Physical Journal D, 19, 49

\bibitem[{Sre\'ckovi\'c {et~al.}(2013)Sre\'ckovi\'c, Mihajlov, Ignjatovi\'c
\& Dimitrijevi\'c}]{sre13}Sre\'ckovi\'c, V.~A., Mihajlov, A.~A., Ignjatovi\'c, L.~M.\& Dimitrijevi\'c, M.~S.
2013, Adv. Space. Res., doi:10.1016/j.asr.2013.11.017.

\bibitem[Stancil et al.(1997)]{sta97} Stancil, P.~C., Kirby,
K., Sannigrahi, A.~B., Buenker, R. J., Hirsch, G., Gu, J.-P.\ 1997, ApJ, 486, 574

\bibitem[{Stilley \& Callaway(1970)}]{sti70}
Stilley, J.~L., \& Callaway, J. 1970, ApJ, 160, 245

\bibitem[Travis
\& Matsushima(1968)]{tra68} Travis, L.~D., \& Matsushima, S.\ 1968, ApJ, 154, 689


\bibitem[{Vernazza {et~al.}(1981)Vernazza, Avrett, \& Loser}]{ver81}
Vernazza, J., Avrett, E., \& Loser, R. 1981, ApJS, 45, 635

\bibitem[{{Ve\v za} {et~al.}(1998){Ve\v za}, {Beuc}, {Milo\v sevi\' c}, \&
  {Pichler}}]{vez98}
{Ve\v za}, D., {Beuc}, R., {Milo\v sevi\' c}, S., \& {Pichler}, G. 1998,
  European Physical Journal D, 2, 45


\bibitem[{Wishart(1979)}]{wis79}
Wishart, A.~W. 1979, MNRAS, 187, 59

\end{thebibliography}

\newcommand{\noopsort}[1]{} \newcommand{\printfirst}[2]{#1}
  \newcommand{\singleletter}[1]{#1} \newcommand{\switchargs}[2]{#2#1}

\appendix

\section{The molecular ion characteristics}

\begin{figure*}
\centering
\includegraphics[height=0.34\textwidth]{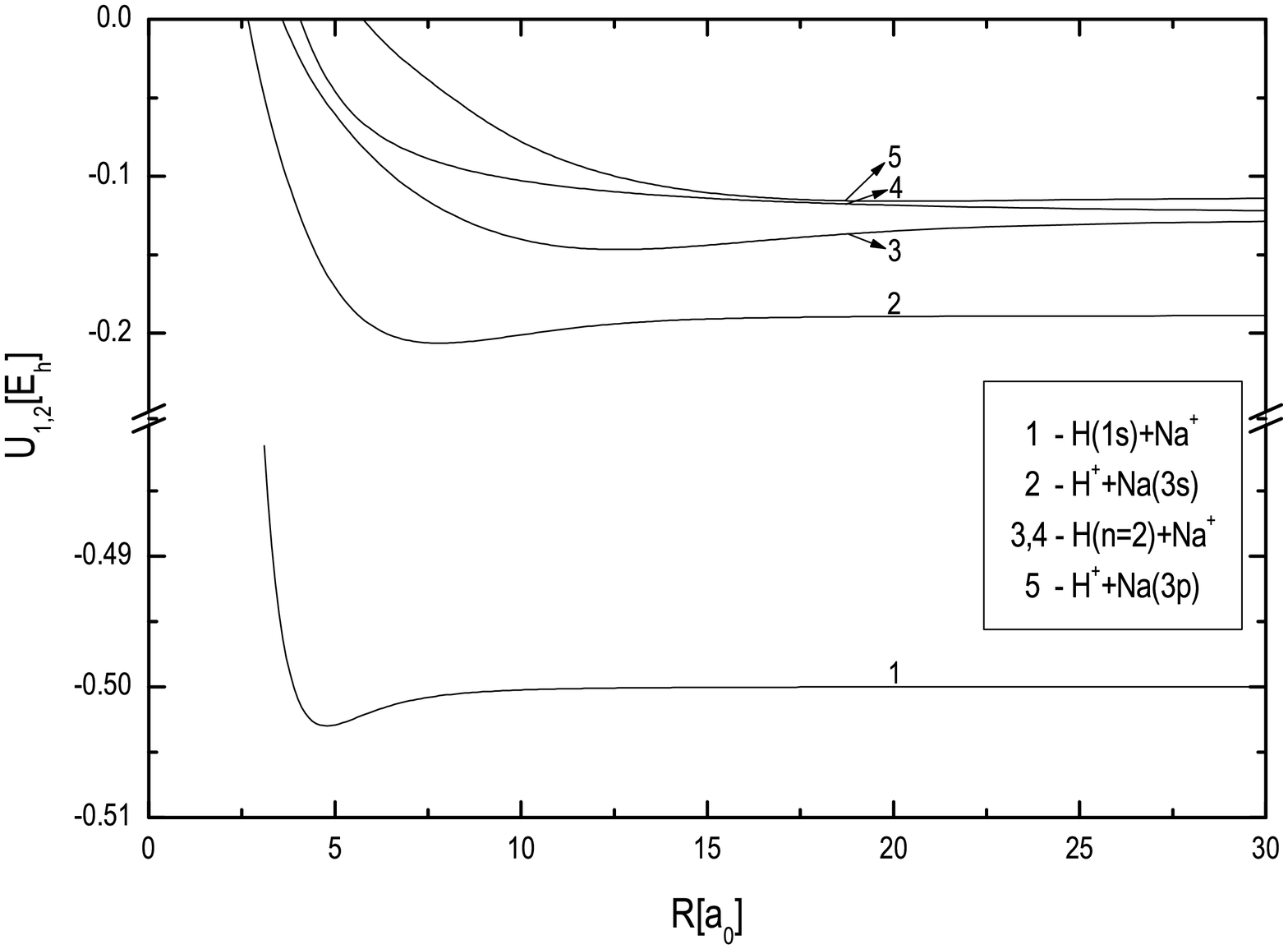}
\includegraphics[height=0.34\textwidth]{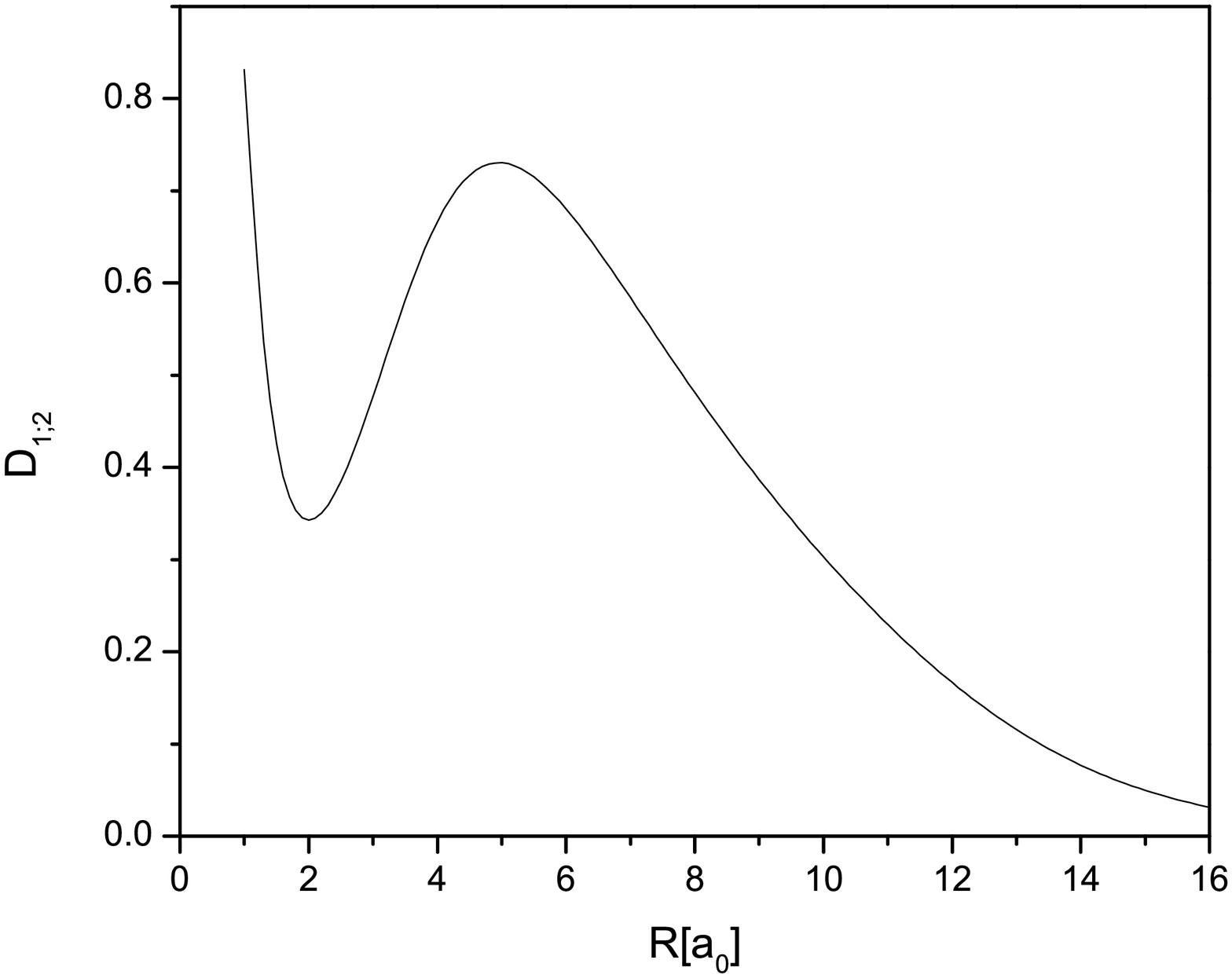}
\caption{\textit{Left panel \bf{a}:} The potential curves of the several low
lying electronic states of the molecular ion HNa$^{+}$.
\textit{Right panel \bf{b}:} The matrix elements of the transition dipole moment for
the given electronic states of the molecular ion HNa$^{+}$.}
\label{fig:Napotdip}
\end{figure*}

\begin{figure*}
\centering
\includegraphics[height=0.34\textwidth]{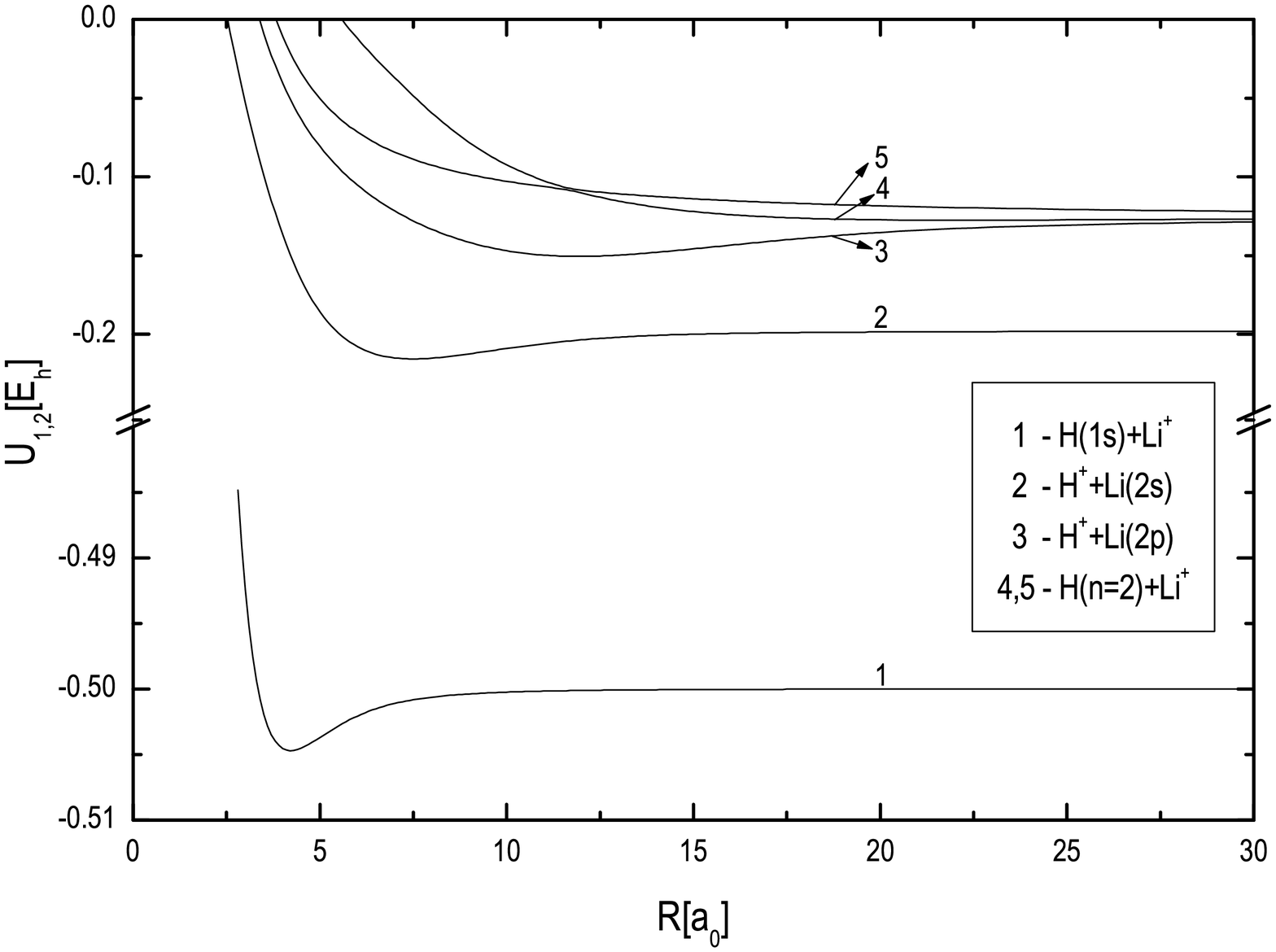}
\includegraphics[height=0.34\textwidth]{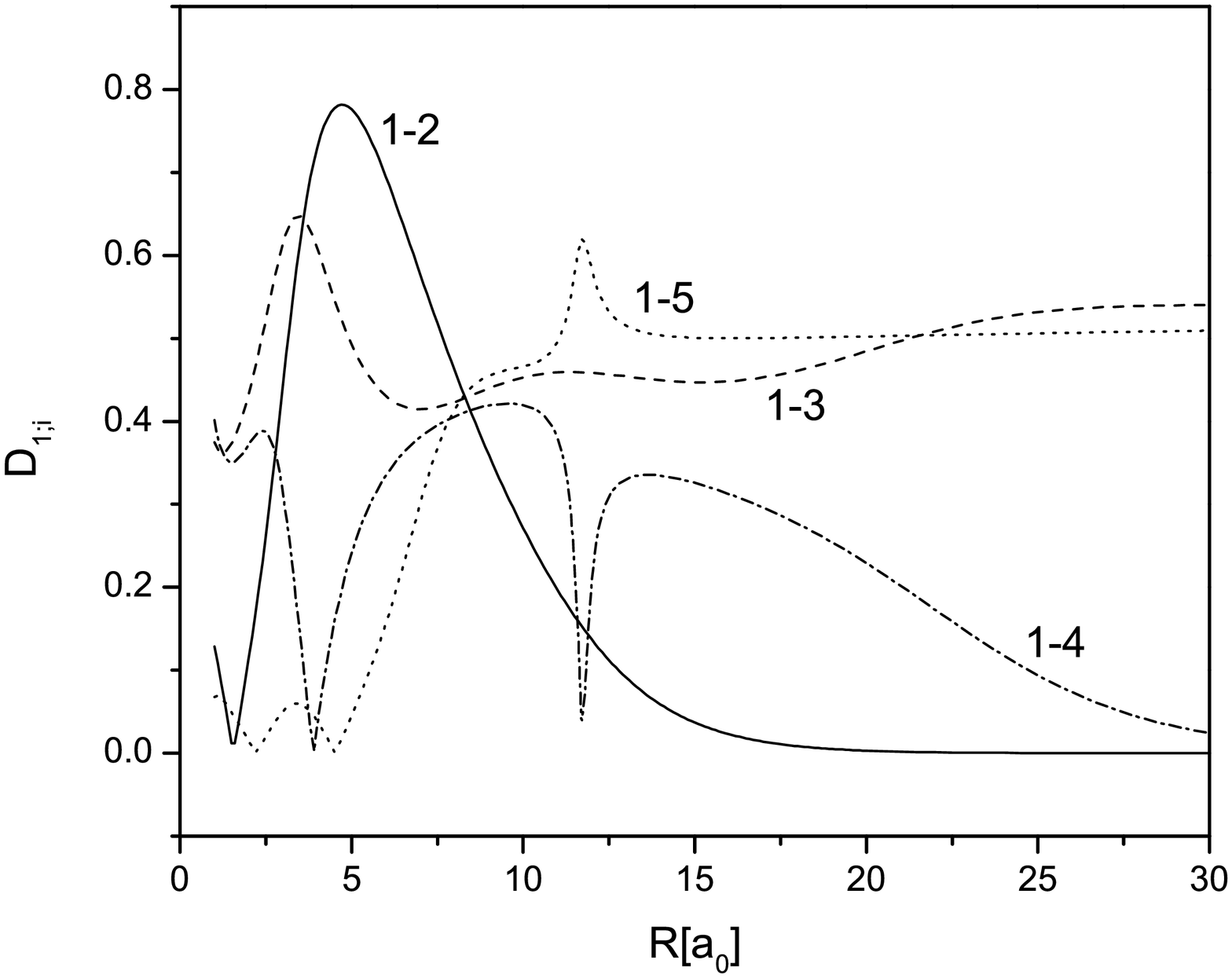}
\caption{\textit{Left panel \bf{a}:} Same as in Fig. \ref{fig:Napotdip}
side a, but for the molecular ion HLi$^{+}$.
\textit{Right panel \bf{b}:} Same as in Fig \ref{fig:Napotdip}
side b, but for the molecular ion HLi$^{+}$.}
\label{fig:Lipotdip}
\end{figure*}

The potential curves of the ground and several low lying excited electronic states
of the molecular ions HNa$^{+}$ and HLi$^{+}$, as well as the corresponding
transition dipole matrix elements, are calculated here by means of the method which
was described in details in \citet{ign05}. The calculated characteristics of these
molecular ions are shown as the functions of the internuclear distance $R$ in
Fig's. \ref{fig:Napotdip} and \ref{fig:Lipotdip}, where the zero of the energy
is chosen in such a way that the potential energy $U_{1}(R)$ of the ground electronic
state is equal to zero at $R = \infty$. In these figures are given the
characteristics not only of the ground and first excited electronic state with the
energy $U_{2}(R)$, but also of the several excited states whose energies are larger then
$U_{2}(R)$. Namely, we keep in mined that such excited electronic state of the
ions HNa$^{+}$ and HLi$^{+}$ could be needed in some further applications.

The mentioned calculation method was developed for the molecular ions $A_{2}^{+}$,
$AB^{+}$ and H$B_{2}^{+}$, where $A$ and $B$ are the alkali metal atoms (Li, Na, etc.).
Within this method the wave functions of the adiabatic electronic states of the
considered molecular ion are described in the single-electron approximation, under
the condition which are analogous to the orthogonality conditions in the known
pseudopotential method of \citet{hei70}. Using this method, in \citet{ign05} were
successfully determined the potential curves and the dipole matrix elements of the
ions Na$_{2}^{+}$ and Li$_{2}^{+}$, and later in \citet{ign08} - of the ion
LiNa$^{+}$. Let us note that the first version of this method was used in
\citet{mih96} just in the connection with the radiative processes in (H +
Li$^{+}$)-collisions.

\section{ The behavior of the spectral characteristics}

\begin{figure*}
\centering
\includegraphics[height=0.34\textwidth]{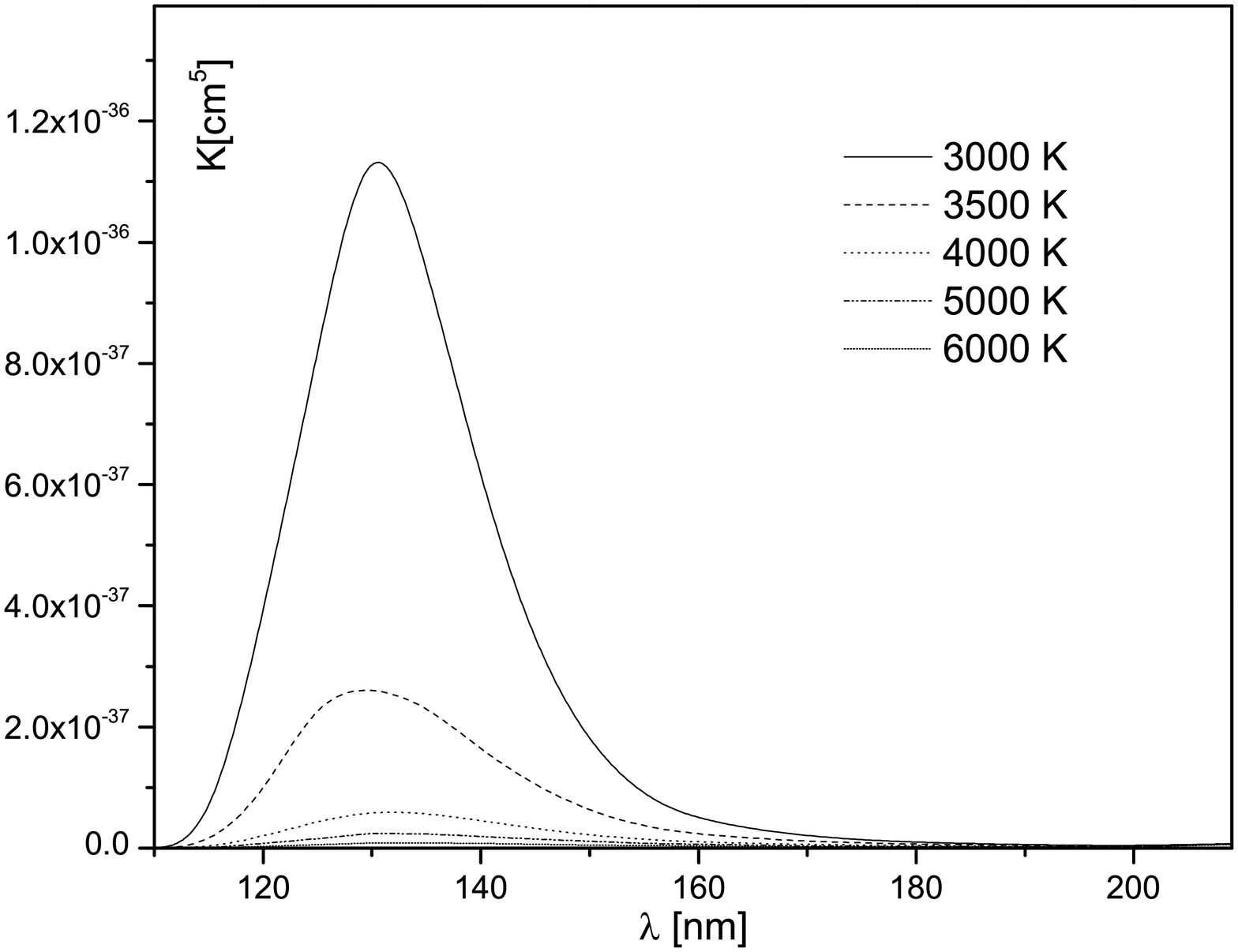}
\includegraphics[height=0.34\textwidth]{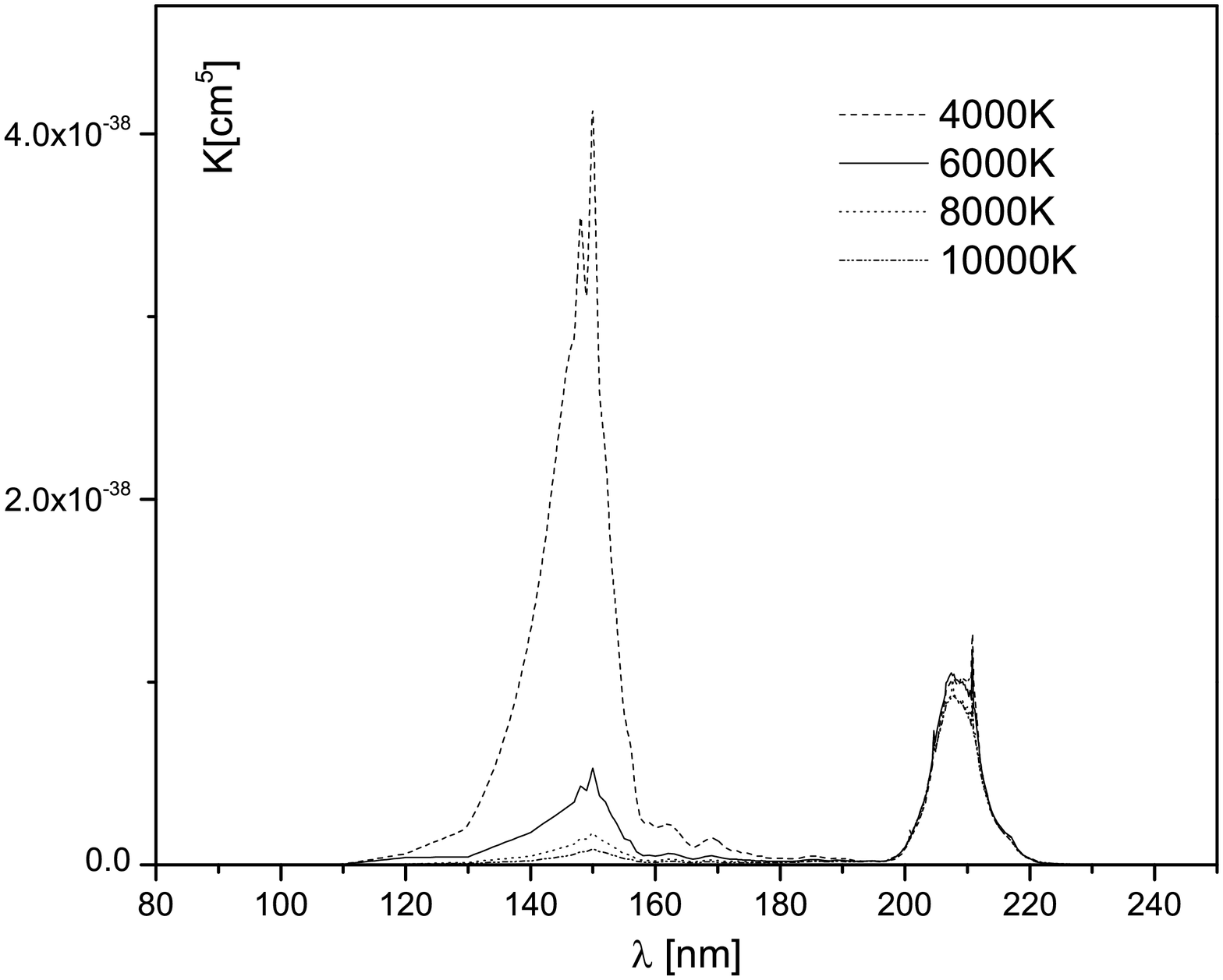}
\caption{\textit{Left panel \bf{a}:} The behavior of the spectral rate
coefficients $K_{X;j=2}(\lambda,T)$, given by  Eq. (\ref{eq:kapansimX}),
which characterize all non-symmetric ion-atom absorption processes
(\ref{eq:nonsim1})-(\ref{eq:nonsim3}) with $X$=Mg.
\textit{Right panel \bf{b}:} Same as in left panel,
but for $X$=Si.}
\label{fig:KMgSi}
\end{figure*}

\begin{figure*}
\centering
\includegraphics[height=0.34\textwidth]{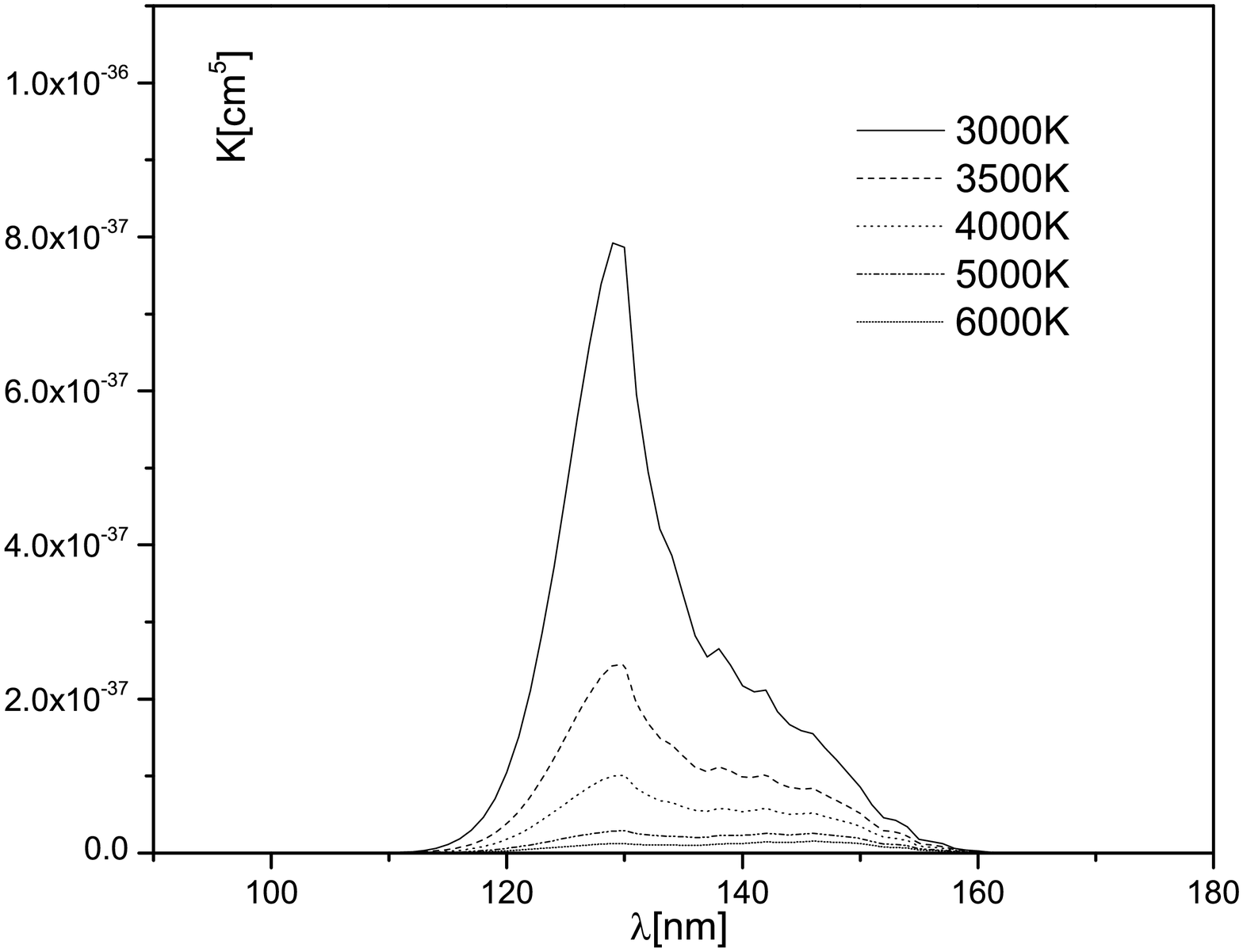}
\includegraphics[height=0.34\textwidth]{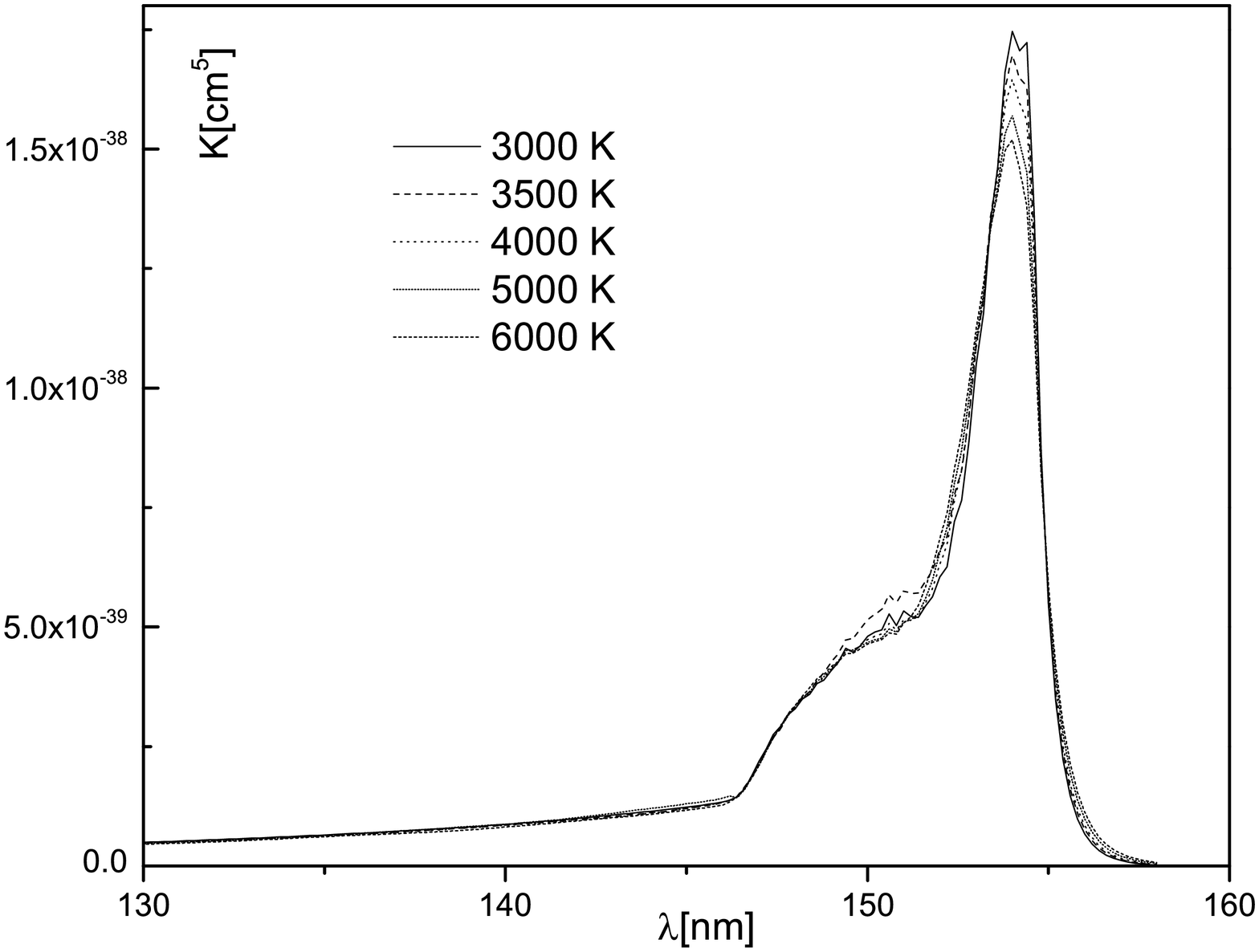}
\caption{\textit{Left panel \bf{a}:} Same as in Fig. \ref{fig:KMgSi},
but for $X$=Ca.
\textit{Right panel \bf{b}:} Same as in Fig. \ref{fig:KMgSi}, but for $X$=Na.}
\label{fig:KCaNa}
\end{figure*}

\begin{figure*}
\centering
\includegraphics[height=0.34\textwidth]{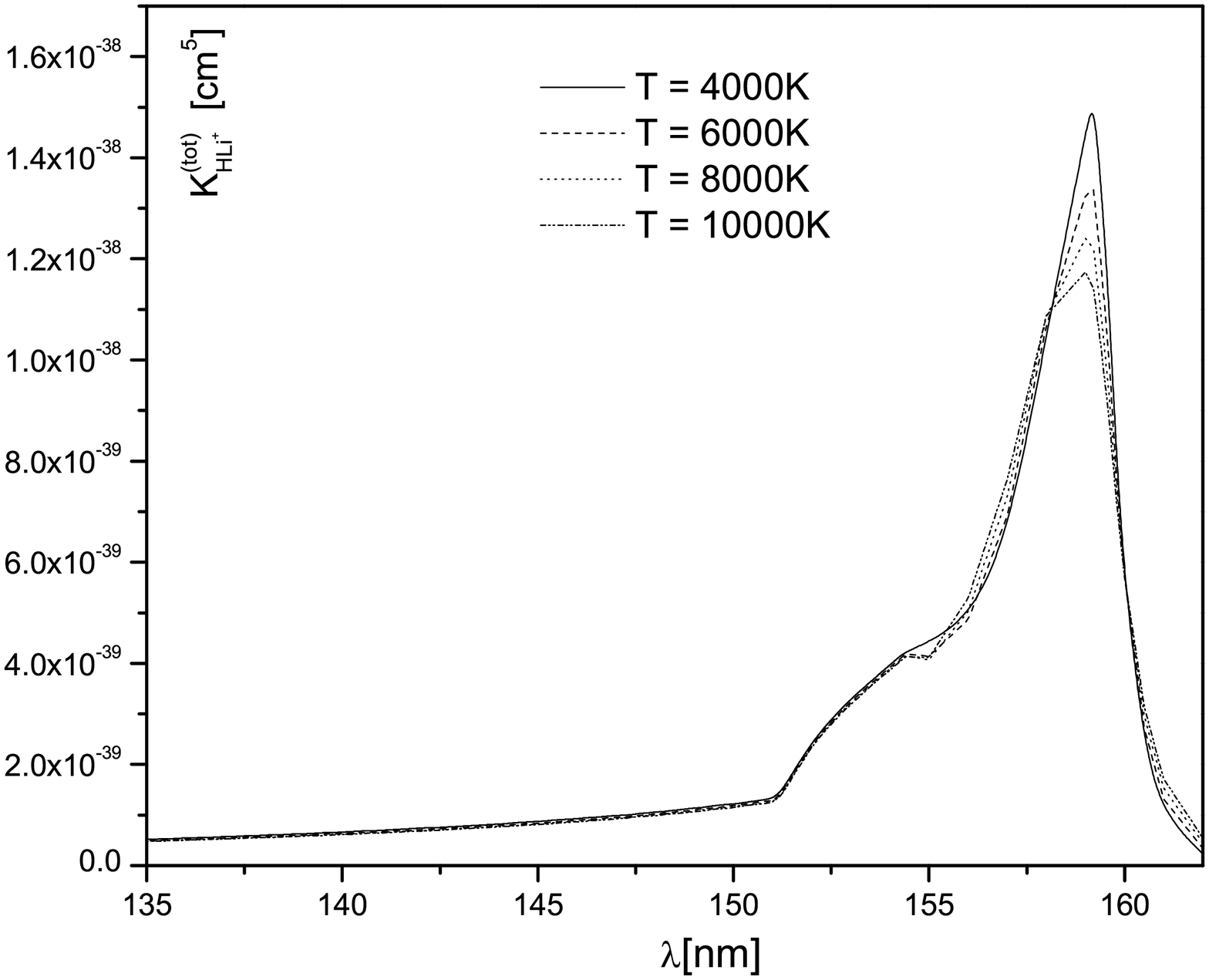}
\includegraphics[height=0.34\textwidth]{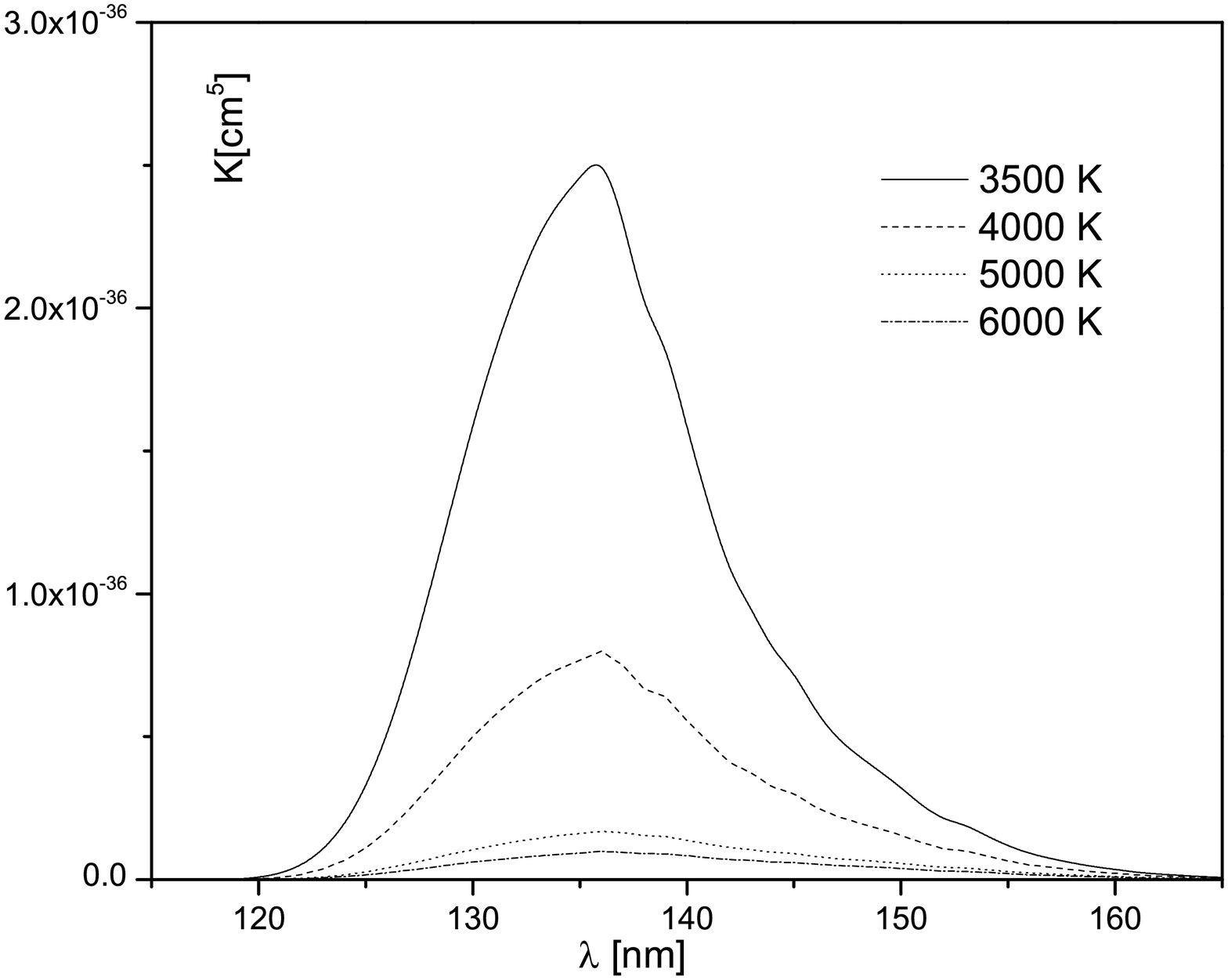}
\caption{\textit{Left panel \bf{a}:} Same as in Fig \ref{fig:KMgSi} but for $X$=Li.
\textit{Right panel \bf{b}:} The behavior of the spectral rate
coefficients $K_{X;j=3}(\lambda,T)$, given by  Eq. (\ref{eq:kapansimX}),
which characterize satellite ion-atom non-symmetric absorption processes
(\ref{eq:sat1})-(\ref{eq:sat3}) with $X^{+*}$=Ca$^{+*}$.}
\label{fig:KLiCa}
\end{figure*}

The spectral rate coefficients $K_{X;2}(\lambda;T)$ and $K_{X;3}(\lambda;T)$ for the
non-symmetric processes (\ref{eq:nonsim1}) - (\ref{eq:nonsim2}) and
(\ref{eq:sat1}) - (\ref{eq:sat2}) respectively, which can be used in some further
applications, are presented here in the relevant regions of $\lambda$ and $T$. Since
the properties of the behavior of the similar spectral rate coefficients were
already discussed (with quite a lot of details) in \citet{mih13}, here they are
shown in Fig's. \ref{fig:KMgSi} - \ref{fig:KLiCa} without additional comments.

\label{lastpage}

\end{document}